\documentstyle[12pt]{article}

\def\hybrid{\topmargin -20pt  \oddsidemargin 0pt
      \headheight 0pt   \headsep 0pt
      \textwidth 6.25in 
      \textheight 9.5in 
      \marginparwidth .875in
      \parskip 5pt plus 1pt   \jot = 1.5ex}

\hybrid

\begin{document}
\def\x{\times}
\def\beq{\begin{equation}}
\def\eeq{\end{equation}}
\def\beqa{\begin{eqnarray}}
\def\eeqa{\end{eqnarray}}
\def\L{ {\cal L}}
\def\C{ {\cal C}}
\def\N{ {\cal N}}
\def\calE{{\cal E}}
\def\lin{{\rm lin}}
\def\Tr{{\rm Tr}}
\def\cF{{\cal F}}
\def\cD{{\cal D}}
\def\modS{{S+\bar S}}
\def\mods{{s+\bar s}}
\newcommand{\Fg}[1]{{F}^{({#1})}}
\newcommand{\cFg}[1]{{\cal F}^{({#1})}}
\newcommand{\cFgc}[1]{{\cal F}^{({#1})\,{\rm cov}}}
\newcommand{\Fgc}[1]{{F}^{({#1})\,{\rm cov}}}
\def\mpl{m_{\rm Planck}}
\def\mxth{\mathsurround=0pt }
\def\xversim#1#2{\lower2.pt\vbox{\baselineskip0pt \lineskip-.5pt
x  \ialign{$\mxth#1\hfil##\hfil$\crcr#2\crcr\sim\crcr}}}
\def\simgr{\mathrel{\mathpalette\xversim >}}
\def\simle{\mathrel{\mathpalette\xversim <}}

\newcommand{\ms}[1]{\mbox{\scriptsize #1}}
\renewcommand{\a}{\alpha}
\renewcommand{\b}{\beta}
\renewcommand{\c}{\gamma}
\renewcommand{\d}{\delta}
\newcommand{\th}{\theta}
\newcommand{\TH}{\Theta}
\newcommand{\pa}{\partial}
\newcommand{\g}{\gamma}
\newcommand{\G}{\Gamma}
\newcommand{\A}{\Alpha}
\newcommand{\B}{\Beta}
\newcommand{\D}{\Delta}
\newcommand{\e}{\epsilon}
\newcommand{\E}{\Epsilon}
\newcommand{\z}{\zeta}
\newcommand{\Z}{\Zeta}
\newcommand{\k}{\kappa}
\newcommand{\K}{\Kappa}
\renewcommand{\l}{\lambda}
\renewcommand{\L}{\Lambda}
\newcommand{\m}{\mu}
\newcommand{\M}{\Mu}
\newcommand{\n}{\nu}
\newcommand{\X}{\Chi}
\newcommand{\R}{\Rho}
\newcommand{\s}{\sigma}
\renewcommand{\S}{\Sigma}
\renewcommand{\t}{\tau}
\newcommand{\T}{\Tau}
\newcommand{\y}{\upsilon}
\newcommand{\Y}{\upsilon}
\renewcommand{\o}{\omega}
\newcommand{\q}{\theta}
\newcommand{\h}{\eta}

\def\dota{ {\dot{\alpha}} }
\def\lag{Lagrangian}
\def\Kahler{K\"{a}hler}
\def\kahler{K\"{a}hler}
\def\A{ {\cal A}}
\def\C{ {\cal C}}
\def\F{{\cal F}}
\def\cL{ {\cal L}}

\def\R{ {\cal R}}
\def\x{ \times }
\def\beq{\begin{equation}}
\def\eeq{\end{equation}}
\def\beqa{\begin{eqnarray}}
\def\eeqa{\end{eqnarray}}

\sloppy
\newcommand{\be}{\begin{equation}}
\newcommand{\eq}{\end{equation}}
\newcommand{\ov}{\overline}
\newcommand{\un}{\underline}
\newcommand{\p}{\partial}
\newcommand{\la}{\langle}
\newcommand{\ra}{\rangle}
\newcommand{\bl}{\boldmath}
\newcommand{\ds}{\displaystyle}
\newcommand{\nl}{\newline}
\newcommand{\Nzahl}{{\bf N}  }
\newcommand{\zzahl}{ {\bf Z} }
\newcommand{\Zzahl}{ {\bf Z} }
\newcommand{\Qzahl}{ {\bf Q}  }
\newcommand{\Rzahl}{ {\bf R} }
\newcommand{\Czahl}{ {\bf C} }
\newcommand{\wt}{\widetilde}
\newcommand{\wh}{\widehat}
\newcommand{\fs}[1]{\mbox{\scriptsize \bf #1}}
\newcommand{\ft}[1]{\mbox{\tiny \bf #1}}
\newtheorem{satz}{Satz}[section]
\newenvironment{Satz}{\begin{satz} \sf}{\end{satz}}
\newtheorem{definition}{Definition}[section]
\newenvironment{Definition}{\begin{definition} \rm}{\end{definition}}
\newtheorem{bem}{Bemerkung}
\newenvironment{Bem}{\begin{bem} \rm}{\end{bem}}
\newtheorem{bsp}{Beispiel}
\newenvironment{Bsp}{\begin{bsp} \rm}{\end{bsp}}
\renewcommand{\arraystretch}{1.5}



\renewcommand{\thesection}{\arabic{section}}
\renewcommand{\theequation}{\thesection.\arabic{equation}}

\parindent0em

\def\S4{\frac{SO(4,2)}{SO(4) \otimes SO(2)}}
\def\P3{\frac{SO(3,2)}{SO(3) \otimes SO(2)}}
\def\MGd{\frac{SO(r,p)}{SO(r) \otimes SO(p)}}
\def\SOd{\frac{SO(r,2)}{SO(r) \otimes SO(2)}}
\def\SO2{\frac{SO(2,2)}{SO(2) \otimes SO(2)}}
\def\SUm{\frac{SU(n,m)}{SU(n) \otimes SU(m) \otimes U(1)}}
\def\SUS{\frac{SU(n,1)}{SU(n) \otimes U(1)}}
\def\SK{\frac{SU(2,1)}{SU(2) \otimes U(1)}}
\def\SU{\frac{ SU(1,1)}{U(1)}}

\begin{titlepage}
\begin{center}
\hfill CERN-TH/96-182\\
\hfill HUB-EP-96/27\\
\hfill SNUTP-96/075\\
\hfill THU-96/25\\
\hfill {\tt hep-th/9607184}\\

\vskip .1in

{\bf HIGHER-ORDER GRAVITATIONAL
 COUPLINGS AND MODULAR FORMS IN $N=2,D=4$
HETEROTIC STRING COMPACTIFICATIONS}

\vskip .2in

{\bf Bernard de Wit$^a$,
Gabriel Lopes Cardoso$^b$, 
Dieter L\"ust$^c$, Thomas Mohaupt$^c$, 
Soo-Jong Rey$^d$}\footnote{email: \tt bdewit@fys.ruu.nl,
cardoso@surya11.cern.ch,
		luest@qft1.physik.hu-berlin.de,\\
	mohaupt@qft2.physik.hu-berlin.de, sjrey@phyb.snu.ac.kr}
\\
\vskip 1.2cm

$^a${\em Institute for Theoretical Physics, 
Utrecht University, 3508 TA Utrecht, Netherlands}\\
$^b${\em Theory Division, CERN, CH-1211 Geneva 23, Switzerland}\\
$^c${\em Humboldt-Universit\"at zu Berlin,
Institut f\"ur Physik, 
D-10115 Berlin, Germany}\\
$^d${\em Physics Department \& Center for Theoretical Physics, 
Seoul National University, Seoul 151-742, Korea}

\vskip .1in

\end{center}

\vskip .2in

\begin{center} {\bf ABSTRACT } \end{center}
\begin{quotation}\noindent
The restrictions of target--space duality are imposed at the 
perturbative level on the holomorphic Wilsonian couplings that 
encode certain higher-order gravitational interactions in 
$N=2, D=4$ heterotic string compactifications. 
A crucial role is played
by non-holomorphic corrections. 
The requirement of symplectic covariance and an associated 
symplectic anomaly equation play an important role in determining their form.
For models which also admit
a type-II description, this equation coincides with
the holomorphic anomaly equation for type-II compactifications in the 
limit that a specific K\"ahler-class modulus grows large. 
We explicitly evaluate some of the higher-order 
couplings for a toroidal compactification with two moduli 
$T$ and $U$, and we  
express them in terms of modular forms.

\end{quotation}
July 1996\\
\hfill CERN-TH/96-182\\
\end{titlepage}
\vfill
\eject

\newpage

\section{Introduction}

Recently,  substantial progress has been achieved
in establishing various types of strong--weak coupling duality
symmetries in superstring theories, such as $S$-duality
of the four-dimensional $N=4$
heterotic string \cite{sduality,SchSen,Sen}
and string--string dualities between
the heterotic and type-II strings \cite{HullTown,Wit1,KV}. 
In fact, it now seems that many of the 
known superstring theories are related to each other either
by non-perturbative
strong--weak coupling duality and/or by perturbative target--space
duality.  It seems that all these different string dualities
can be understood in a unified manner in the 
new framework of either $M$-theory \cite{Wit1,Schwarz}
or $F$-theory \cite{ftheory}.

One interesting case of string--string duality is 
the duality between heterotic and type-II strings with $N=2$ 
supersymmetry in four space--time
dimensions.  Such theories exhibit a non-perturbative structure 
which, in the point--particle limit, contain  \cite{KKLMV}
the non-perturbative effects of rigid
$N=2$ supersymmetric field theories
studied in \cite{SW}. A large class of $N=2$ heterotic vacua in four
dimensions can be obtained by the compactication of the ten-dimensional
heterotic string on $K_3\times T_2$. The corresponding $N=2$ type-II vacua
are constructed by compactifying the ten-dimensional type-II string
on suitably chosen Calabi--Yau three-folds.  
This string--string duality has been tested in 
several models with a small
number of vector supermultiplets, and it 
has successfully passed various non-trivial explicit checks
\cite{KV},\cite{KLM}--\cite{CCLM}. Most of these
tests were based on the comparison of lower-order gauge
and gravitational couplings \cite{DWKLL,AFGNT,CLM}
of the perturbative heterotic string 
with the corresponding couplings of the
dual type-II string in some corner of the Calabi--Yau moduli space.
That is, it was shown that the perturbative heterotic prepotential
${\cal F}^{(0)}$ and the function $\cFg{1}$ (which specifies the 
non-minimal gravitational interactions 
involving the square of the Riemann tensor) 
agree with the corresponding type-II functions
in the limit where one specific K\"ahler-class modulus of the
underlying Calabi--Yau space becomes large.  A set of interesting
relations between certain topological Calabi--Yau data (such as
intersection numbers,
rational and elliptic instanton numbers) and various modular forms has
emerged when performing these tests \cite{HarvMoo,CCLM}.

In order to show that a  given $N=2$ string model has two dual 
descriptions, both a heterotic and
a type-II description, it is important to also verify whether 
string--string duality holds for higher-order (nonminimal) 
gravitational couplings.
A particularly interesting subset of such couplings is based on 
chiral $N=2$ densities which involve the square of the Riemann tensor 
multiplied by powers of the graviphoton field strength. These invariants 
take the form ${\cal F}^{(g)}\,R^2\,T^{2(g-1)}$, 
where $T$ is related to the $N=2$ graviphoton
field strength in a way we will specify later, 
and ${\cal F}^{(g)}$ is a function which depends on the vector
moduli.  For type-II string compactifications, 
it was shown in \cite{AGNT1} that
these higher-order couplings satisfy a holomorphic
anomaly equation derived for the topological genus-$g$ partition
functions of twisted Calabi--Yau sigma models \cite{BCOV}.  
If a given type-II model is
to have a dual heterotic description, then the heterotic higher-order 
gravitational 
couplings should satisfy similar holomorphic anomaly equations.
For the case of a particular model with gauge group $U(1)^3$, 
the so called $S$-$T$ model,
it was shown in \cite{AGNT2} that some of the heterotic
higher-order couplings indeed satisfy the anomaly equations of 
\cite{BCOV}, at least at the perturbative level.

On the other hand, target--space duality symmetry is a manifest
symmetry at weak string coupling in heterotic string compactifications.
Hence, this symmetry should be encoded in the perturbative
heterotic prepotential as well as in all the gravitational couplings
${\cal F}^{(g)}$.  These target--space
duality transformations constitute a subgroup of the $N=2$
symplectic reparametrizations. 
However, the (holomorphic) Wilsonian couplings $\cFg{g}$ do not 
correspond directly to physical quantities and therefore are not 
themselves invariant under target--space duality transformations. In fact, 
it turns out that the holomorphic couplings do not transform
as functions (or rather, sections) under symplectic reparametrizations
and non-holomorphic terms are necessary in order to obtain quantities 
that do transform in a covariant form \cite{DW}. For these quantities one
can then easily formulate the requirement of target--space duality
invariance, which, as it turns out, can be translated into 
certain complicated restrictions on the original holomorphic functions. 
A particular proposal for the minimal non-holomorphic corrections required 
for symplectic covariance, was presented in the second paper of 
\cite{DW}, where it should be noted that this construction does not 
exclude the possibility of additional non-holomorphic
terms as long as they constitute an independent symplectic 
function. The result 
turns out to satisfy a certain holomorphic anomaly equation, which
henceforth will be referred to as the `symplectic' anomaly equation 
in order to distinguish it from the anomaly equation of \cite{BCOV}.

As alluded to above, holomorphic anomaly equations can be derived
in the context of topological field theories. They can also be understood 
in a space--time context as resulting from the propagation of massless 
modes. 
For those heterotic
$N=2$ models admitting a type-II description, 
we can make use of string--string
duality 
and consider the anomaly equation of \cite{BCOV}
in the limit where one of the type-II K\"ahler-class  moduli is taken 
to be large so as to make contact with the perturbative heterotic description. 
Interestingly, 	
 in this limit the anomaly equation of \cite{BCOV} coincides with 
the symplectic anomaly equation of \cite{DW}.
We further demonstrate that, in the heterotic weak-coupling limit,
this anomaly equation is consistent with
target--space
duality transformations.  In doing so, one has to take into account
that, at the one-loop level, 
the dilaton field is no longer invariant 
under target--space duality
 transformations and neither is the so-called 
Green--Schwarz term (describing the mixing of the dilaton field with 
the moduli), which also appears in the anomaly equation. 
We show that these two effects 
compensate each other, and by reformulating everything in terms
of the invariant dilaton 
field and the invariant Green--Schwarz term \cite{DWKLL},
the results become manifestly covariant under target--space duality
transformations.

In order to elucidate the above observations, we will consider the 
so-called
$S$-$T$-$U$ model in detail. This is a heterotic rank-4 model,
that is, a model with gauge group $U(1)^4$,
which is believed to have a dual type-II interpretation \cite{KV}. 
We will
solve the 
relevant anomaly equation in the heterotic weak-coupling limit
for the higher-order gravitational couplings $\cFgc{2}$ and 
$\cFgc{3}$.  We will show that  
the results for the covariant non-holomorphic couplings
$\cFgc{2}$ and 
$\cFgc{3}$ can be cast in a form that is explicitly covariant under 
target--space duality transformations by expressing them in terms of 
modular forms.
In general, when solving for the non-holomorphic couplings $\cFgc{g}$,
one encounters holomorphic ambiguities \cite{BCOV}, which cannot be 
fixed unless further inputs are provided.  Interestingly, for the 
$S$-$T$-$U$ model, 
$\cFgc{2}$ is free from these holomorphic ambiguities, because 
target--space duality covariance and the 
knowledge of the leading holomorphic
singularities, that are associated with known 
gauge--symmetry enhancement points/lines,  fixes its structure 
completely.  An unambiguous determination of $\cFgc{2}$ provides
one with information of genus-2 instanton numbers
for the corresponding dual Calabi--Yau model $WP_{1,1,2,8,12}(24)$.
This is an example of the utility of the second-quantized 
mirror map \cite{FerHarStrVa}.  If the higher $\cFgc{g}$ could also
be determined unambiguously, then one would
in principle be able to 
obtain information about the higher-genus
instanton numbers on the dual type-II side, about which not much is
known.

This paper is organized as follows. In section 2 we
will discuss the projective transformation properties of the higher-order
 couplings ${\cal F}^{(g)}$.  In section 3 we 
discuss their behaviour under symplectic reparametrizations and 
extend the
discussion of 
the symplectic anomaly equation given in \cite{DW}.
We will also provide the explicit solutions to the
symplectic anomaly equation for a specific class of $N=2$
effective field theories based on a cubic prepotential.
In section 4 we test string--string duality by showing that
the symplectic and the holomorphic anomaly
equations agree with each other in the heterotic weak-coupling limit.
We then solve
the holomorphic anomaly equations for the heterotic higher-order
couplings
${\cal F}^{(2)}$
and ${\cal F}^{(3)}$ in the $S$-$T$-$U$ model.
When solving the anomaly equation for ${\cal F}^{(2)}$, we will properly
take into account the Green--Schwarz term which arises as a dilaton--moduli
mixing term in the one-loop prepotential. 
We present our conclusions
in section 5.  We refer to the various appendices for additional
information on 
some of the more technical aspects of our calculations.


\section{The holomorphic sections $\cFg{g}(z)$}
\setcounter{equation}{0}

Consider an $N=2$ supersymmetric effective field theory based on vector 
supermultiplets with generic couplings to supergravity, both of
the `minimal' and the `nonminimal' type. The latter incorporate
interactions of vector multiplets with the square of the Riemann
tensor and are based on the Weyl multiplet:  
an $N=2$ (reduced) chiral superfield $W^{ij}_{ab}$, which comprises the 
covariant quantities of the conformal supergravity sector.%
\footnote{%
   The reduction of the Weyl multiplet is implemented by the
   following (linearized) constraint,
   ${\rm Im}\; \big(\bar D_i\s^{ab} D_j\, W_{ab}^{ij}\big)=
   0$, where $i,j$ are chiral $SU(2)$ indices and $a,b$ denote
   Lorentz indices \cite{BDRDW}.} %
This superfield is antisymmetric in both types of 
indices and anti-selfdual as a Lorentz tensor. 
Besides the graviton and gravitino field strenghts, the covariant 
quantities consist of the field strengths of the gauge fields 
associated with the chiral $SU(2)\times U(1)$ automorphism group 
of the $N=2$ supersymmetry algebra, and an auxiliary spinor, scalar 
and tensor field. The anti-selfdual part of the latter, denoted 
by $T^{ij}_{ab}$, equals the lowest-$\theta$ component of 
$W^{ij}_{ab}$. The Riemann tensor resides at the 
$\theta^2$-level, modified by terms proportional to $T_{ab\,ij}\,
T^{ij}_{cd}$, as well as the $SU(2)\times U(1)$ field 
strengths. The highest-$\theta$ term contains second derivatives of 
$T_{ab\,ij}$ \cite{BDRDW}.  

It is instructive to compare the Weyl multiplet to the vector 
supermultiplet. The covariant quantities of the latter constitute 
a reduced (scalar) chiral field (i.e., the superfield strength), 
whose lowest-$\theta$ component is the complex scalar of the 
vector multiplet.%
\footnote{%
   For the superfield strength $\Phi$ the reduction is effected
   by the superspace constraint 
   ${\rm Im}\; \big(\bar D_{i} D_{j}\, \Phi\big)= 0$. } %
We denote these scalars by $X^I$, where 
the index $I$ labels the various vector multiplets ($I=0, 1,
\ldots,n$). Supergravity couplings of these multiplets depend
sensitively on their assignment under dilatations and 
$U(1)$ transformations. As it turns out $T^{ij}_{ab}$ and 
the scalar fields $X^I$ share the same dilatational and $U(1)$ 
weights, equal to $+1$ and $-1$, respectively. 
Obviously, their complex conjugates, the selfdual tensor $T_{ab\,
ij}$ and the scalars $\bar X^I$, carry opposite $U(1)$ 
weights. 

{}From the Weyl multiplet one constructs a scalar chiral multiplet 
of weight 2, by taking its square $W^2\equiv 
(W^{ij}_{ab}\varepsilon_{ij})^2$. Owing to its tensorial structure
no other independent products of $W$ can appear in a chiral
scalar density. The 
general `nonminimal' coupling of vector multiplets and
supergravity is now 
encoded in a holomorphic function of the $X^I$ and $W^2$. A 
consistent coupling to supergravity requires this function to
be homogeneous of second degree \cite{SU},
\beq
F(\l X,\l^2W^2) = \l^2 F(X,W^2)\,. \label{holofunction}
\eeq
One may expand $F$ in powers of $W^2$ and write
\beq
F(X,W^2) = \sum_{g=0}^\infty \Fg{g}(X) \, \big[W^2\big]^g
\,. \label{holoexpansion}
\eeq
The coefficient functions $\Fg{g}$ are holomorphic homogeneous 
functions of the $X^I$ of degree $2-2g$. In this paper we intend
to study the implication of target space duality invariance for these
quantities in certain string compactifications. 

The function $\Fg{0}$, which is thus homogeneous of second degree, 
determines the self-interactions of the vector  
supermultiplets with `minimal' coupling to supergravity. 
Henceforth we drop the superscript $(0)$ and simply write $F(X)$; to 
avoid confusion we will be careful and indicate the arguments $X$ 
and $W^2$ whenever referring to the full function $F(X,W^2)$. 
Initially the action takes a form 
that is invariant under local  
dilatations. As a result of this, the coefficient of the 
Ricci scalar contains a field-dependent factor 
proportional to $i (\bar X^I\,F_I - X^I\,\bar F_I)$.%
\footnote{%
   We use the standard notation where $F_{IJ\cdots}$ denote 
   multiple derivatives of $F(X)$ with respect to $X$.} %
Without loss of generality we can apply a local dilatation to set 
this coefficient equal to the Planck mass such as to obtain the
Einstein--Hilbert Lagrangian. Hereby the scalar fields will be 
restricted to an $n$-dimensional complex hypersurface. It is then 
convenient to parametrize the scalars in terms of holomorphic 
sections $X^I(z)$ depending on $n$ complex coordinates $z^A$, 
which describe the hypersurface projectively. In terms of these 
sections the $X^I$ read
\beq
X^I =\mpl \, e^{{1\over 2}K(z,\bar z)}\,
X^I(z)\,.\label{section} 
\eeq
In order to distinguish the sections $X^I(z)$ from the original quantities
$X^I$, we will always explicitly indicate their $z$-dependence.
The overall factor $\exp[{1\over2}K]$ is chosen such that $i 
(\bar X^I\,F_I - X^I\,\bar F_I)=\mpl^2$. With this
requirement $K(z,\bar z)$ equals 
\begin{equation}
K(z,\bar z)=
-\log\Big(i \bar X^I(\bar z)\,F_I(X(z)) -i X^I(z)\,
\bar F_I(\bar X(\bar z))\Big) \,,
\label{KP}
\end{equation}
and coincides with the K\"ahler potential associated with the 
target--space metric for the complex fields $z^A$. 

As mentioned above, the sections are defined projectively, 
i.e., modulo multiplication by an arbitrary holomorphic 
function,
\beq
X^I(z)\longrightarrow e^{f(z)}\, X^I(z)\,. \label{projtran}
\eeq
These projective transformations induce corresponding K\"ahler
transformations on the  K\"ahler potential,
\beq
K(z,\bar z)\longrightarrow K(z,\bar z) -f(z) -\bar f(\bar z)\,. 
\label{Ktrans}
\eeq
On the original quantities $X^I$ the transformation 
(\ref{projtran}) induces a phase transformation. This $U(1)$ 
transformation acts on all quantities that carry nonzero chiral weight. 
Obviously, the consistency of the above formulation depends on
the presence of the aforementioned local $U(1)$. 

Eventually one has to fix the parametrization of the holomorphic 
sections (i.e., impose a gauge), after which the freedom to
perform the transformations 
(\ref{projtran}) disappears. A convenient way 
to do this, is by choosing so-called {\it special} coordinates
corresponding to  
$X^0(z)=1$ and $X^A(z)=z^A$. In that case we have $|X^0|^2 =\mpl^2 \exp 
\big[K(z,\bar z)\big]$. In the context of a specific holomorphic
parametrization, certain 
transformations of the holomorphic sections will be
accompanied by corresponding projective transformations in order
to ensure that one remains within the chosen gauge. 

Without the dependence on $W^2$ the $U(1)$ gauge 
field can be integrated out straigthforwardly, which leaves the 
local invariance intact. With the interactions to $W^2$ the 
integration over auxiliary fields is more subtle and can be done 
iteratively order--by--order in the inverse Planck mass. To 
preserve supersymmetry, elimination of the auxiliary 
fields should be postponed until the end. 
For future reference we give the value of the auxiliary field
$T_{ab}^{ij}$, 
\beq
T_{ab}^{ij}= -{2i\varepsilon^{ij}\over \mpl^2}
\left[X^I\,\bar{\cal N}_{IJ}  F_{ab}^{-J} 
-F_I F_{ab}^{-I}\right] + \cdots\,. \label{Tvalue}
\eeq
Here $\cal N$ is the matrix that appears
in the kinetic terms for the gauge fields, which satisfies ${\cal
N}_{IJ} X^J = F_I$. Note that the first term is of order
$1/\mpl$ by virtue of (\ref{section}); the ellipses
denote fermionic terms and terms that are suppressed by
additional negative powers of 
$\mpl$; the latter arise as a result of the nonminimal
supergravity interactions. Because $T$ is a superconformal
background field, the expression (\ref{Tvalue}) should be 
insensitive to symplectic transformations of the vector
supermultiplets, which we will discuss in due course. This is
indeed the case, because both 
$(X^I,F_J)$ and $(F_{ab}^{-I}, 
\bar{\cal N}_{JK}F_{ab}^{-K})$ transform as symplectic vectors. While 
$F_{ab}^{-I}$ corresponds to the (generalized) electric and 
magnetic induction fields, $\bar{\cal N}_{IJ} F_{ab}^{-J}$
describes the (generalized) electric displacement and magnetic fields.  

Under supersymmetry the gravitino fields do not transform
directly into the 
vector fields, but into the auxiliary field $T$. Therefore, the
field-dependent linear combination of the fields strengths given 
in (\ref{Tvalue}) defines the graviphoton field strength.

Let us now return to the case where interactions with $[W^2]^g$ are
taken into account. For $g>1$ the presence of the vector 
multiplets is crucial in order to compensate for the lack of conformal 
invariance of $[W^2]^g$. The Lagrangian based on the chiral
superspace integral of just $W^2$ is superconformally 
invariant; its full nonlinear component form can be found in 
\cite{BDRDW}. The real part leads to the supersymmetric extension
of the square of the Weyl tensor and should be regarded as the
action of conformal supergravity. Its imaginary part is a total
divergence whose space--time integral leads to a topological
quantity, the Hirzebruch signature.
The action contains the standard gauge--invariant kinetic
terms for the $SU(2)\times U(1)$ gauge fields and a kinetic term
for the tensor field $T^{ij}_{\m\n}$. There is only one (scalar)
field that appears in  
this action as an auxiliary field without derivatives. 

With the interactions to the vector multiplets we introduce
further modifications. The square of the Weyl tensor now acquires
modifications by vector multiplet scalars and the tensor field of
the form $\Fg{g}(X)\,(T^{ij}_{cd}
\varepsilon_{ij})^{2(g-1)}\,R^2$, but there will also be
modifications of the kinetic terms for the vector fields
proportional to $(T^{ij}_{cd}
\varepsilon_{ij})^{2g}\,\Fg{g}_{IJ}(X)\, 
F_{ab}^{-I}F_{ab}^{-J}$.

When substituting (\ref{section}) into the various coefficient
functions, the $F^{(g)}$ are replaced by sections according to 
\beq
\cFg{g}(z) = i\, [\mpl^2]^{g-1}\,e^{-(1-g) K(z,\bar z)}\,\Fg{g}(X)
\,. \label{Fsections} 
\eeq
With these definitions the Einstein--Hilbert action and the
nonlinear sigma model action of the K\"ahler manifold  acquire an
explicit factor $\mpl^2$. The nonminimal interactions that
involve the square of the Riemann tensor are multiplied by the
holomorphic sections
$\cFg{g}$ times an explicit factor $(\mpl^2)^{1-g}$. 

Finally we note that under projective transformations these
sections transform as  
\beq
\cFg{g}(z)\longrightarrow e^{2(1-g)f(z)}
\,\cFg{g}(z)\,. \label{Fprojective} 
\eeq

\section{Symplectic transformations of the $\cFg{g}(z)$ and 
holomorphic anomalies \label{shanom}}
\setcounter{equation}{0}

One of the aims of this paper is to investigate whether we can
constrain the quantities $\cFg{g}$ for certain string
compactifications by imposing the requirement of
target--space duality. This approach turned out to be successful
for the minimal 
supergravity interactions encoded in the $W^2$-independent part
of $F(X, W^2)$, as was described in \cite{DWKLL,AFGNT}. 
Here it is important to appreciate that the target--space duality group is
part of the $Sp(2n+2;{\bf Z})$ group of symplectic
reparametrizations of the vector supermultiplets.%
\footnote{%
   As usual the classical action allows $Sp(2n+2;{\bf R})$
   transformations, but nonperturbatively this group is restricted
   to an integer-valued subgroup.} %
Under these reparametrizations $F(X,W^2)$ does not transform as a
function; 
the new function assigned to it takes a more complicated form, as
we will show below. This implies that invariance requirements
cannot be directly imposed on the sections $\cFg{g}$. As it
turns out this feature is generically intertwined with the
presence of non-holomorhic additions to the $\cFg{g}$
\cite{DW}. These non-holomorphic terms lead to a so-called
holomorphic anomaly \cite{DKL,AGNT1,BCOV}, which can be
understood from the contributions due to massless fields. The
purpose of the text below is to elucidate this. 

On the scalar fields the symplectic transformations act according
to
\beq
\pmatrix{X^I\cr F_I(X,W^2)} \longrightarrow \pmatrix{\tilde
X^I\cr \tilde  F_I(\tilde X,W^2)} = \pmatrix{U& Z\cr W& V}
\pmatrix{X^I\cr F_I(X,W^2)\cr}\,. 
\label{Xtrans} 
\eeq
There are similar transformations on the (abelian) field
strengths $F^{\pm I}_{\m\n}$ and $G^\pm_{\m\n\,I} = {\cal N}_{IJ} 
F^{\pm J}_{\m\n} + {\cal O}_{\m\n\,I}^\pm$ where ${\cal
O}^\pm_{\m\n\,I}$ represents  
moment couplings to the vector fields, such that the 
Bianchi identities and the field equations read  
$\pa^\n(F^+- F^-)_{\m\n}^I = \pa^\n(G^+- G^-)_{\m\n\, I} =0$.
Note that it is crucial here  to include the full dependence on the Weyl
multiplet, also in the tensors $G^\pm_{\m\n\,I}$ and ${\cal
O}^\pm_{\m\n\,I}$. The reason is that the symplectic
transformations are 
linked to invariances of the full equations of motion for the vector
fields (which involve the Weyl multiplet) and not of (parts of)
the Lagrangian. Note that the transformations are $W$--dependent
and holomorphic (both in $X$ and $W$), but that $W$ itself does
not transform under the symplectic transformations.

The matrix in (\ref{Xtrans}) constitutes an element of
$Sp(2n+2;{\bf Z})$.  
The transformation rule (\ref{Xtrans}) specifies the
reparametrization of the fields $X^I\to \tilde X^I$ and express
the change of 
the first derivatives of the function $F(X,W^2)$. Owing to the
symplectic nature of the transformation, the change in the
latter is such that the new quantities $\tilde F_I$ are again the
derivative of some new function, which we denote by $\tilde
F(\tilde X,W^2)$. It is possible to integrate the expression for
the $\tilde F_I$ and determine $\tilde
F(\tilde X,W^2)$, up to certain integration constants. The result
reads 
\beqa
\tilde
F(\tilde X,W^2)&=& F(X,W^2) - {\textstyle{1\over 2}} X^I F_I
\label{Ftilde}  \label{newfunction} \\ 
 && +  {\textstyle{1\over 2}}\Big[(U^{\rm T}W)_{IJ}  X^IX^J +
(U^{\rm T}V+  W^{\rm T}Z)_I{}^JX^IF_J + (Z^{\rm T}V)^{IJ}
F_IF_J\Big] \,, \nonumber  
\eeqa
where $F_I$ denotes the derivative of $F(X,W^2)$ with respect to
$X^I$.  Clearly $F(X,W^2)$ itself does not transform as a
function, although the combination $F(X,W^2)-{1\over 2} X^I
F_I(X,W^2)$ does. The symplectic reparametrization constitutes an
invariance of the equations of motion, if the new function is
identical to the old on, i.e., iff
\beq
\tilde F(\tilde X,W^2)= F(\tilde X, W^2)\,. \label{invariantF}
\eeq
Again, the above equation is {\it not} equivalent to the
requirement that $F(X,W^2)$ is an invariant {\it function}. 

By differentiating (\ref{newfunction}) with respect to $W^2$ and
putting $W^2=0$ at the end, one can derive the 
transformation rules for the coefficient functions
$F^{(g)}(X)$. In this way one establishes that, with
the exception of $g=1$, none of the coefficient
functions transforms as a function. More explicitly, $F^{(1)}$
changes under symplectic reparametrizations according to
\beq
\tilde \Fg{1} (\tilde X) = \Fg{1}(X) \,, \label{holF1}
\eeq
while the $\Fg{g>1}(X)$ transform in a rather complicated way that
involves all lower-$g$ functions as well \cite{DW}. Observe that
we are still formulating the transformation rules for expressions
depending on the $X^I$ rather than on the sections $X^I(z)$,
but we shall turn to this aspect shortly. 

It is possible to introduce modifications to the $\Fg{g}$, such
that they become symplectic 
functions. As it turns out, these modifications are necessarily
non-holomorphic, although their precise form cannot be determined
solely by the requirement of symplectic covariance, because one can
always consider the addition of other symplectic
functions. Hence, there is no contradiction between the
above result, which identifies the  
Wilsonian coefficient function $\Fg{1}$ as both symplectic
and holomorphic, while we know, for instance from string theory, that it
acquires an antiholomorphic contribution. Presumably this simply implies
that an independent symplectic but 
nonholomorphic function must be added. Obviously, for the generic
$\Fg{g}$ the non-holomorphic corrections will always transform into
holomorphic terms, so as to compensate for the previous lack of
symplectic covariance. 

The minimal nonholomorphic modifications that are required to
turn the Wilsonian coefficient functions into symplectic
functions, can be written down systematically by making use of
the following derivative $\cal D$ \cite{DW},
\beq 
{\cal D}\,\equiv\, {\pa\over \pa W^2} +  
i{\pa^2 F(X,W^2)\over \pa W^2\,\pa X^I} \,N^{IJ}(X,\bar
X,W^2,\bar W^2)\,{\pa\over \pa X^J} \,,  \label{derivative}  
\eeq
where
\beq 
N_{IJ}(X,\bar X,W^2,\bar W^2) \equiv 2\,{\rm Im}
\,F_{IJ}(X,W^2)\,, \qquad N^{IJ}\equiv  
\big[N^{-1}\big]^{IJ}\,,  \label{defN} 
\eeq
so that $\cal D$ is non--holomorphic in both $X$ and $W$. 
The derivative (\ref{derivative}) is constructed such that when
acting on a quantity $G(X,\bar X,W^2,\bar W^2)$ that transforms as a
function under symplectic transformations, i.e., as
\beq
\tilde G(\tilde X,\bar{\tilde X},W^2,\bar W^2)= G(X,\bar X,W^2,\bar
W^2) \,,
\eeq 
then also ${\cal D}G(X,\bar X, W^2,\bar W^2)$ will transform as a
symplectic function. Using (\ref{derivative}) one can thus  
write down a hierarchy of functions which are modifications of
the Wilsonian coefficient functions $\Fg{g}$,
\beq
\Fgc{g} (X,\bar X) \,\equiv \, {1\over g!}\,\Bigg[{\cal D}^{g-1}
\,{\pa F(X,W^2)\over \pa W^2} \Bigg]_{W^2 =0}\, ,\label{Fg}
\eeq
where we included an obvious normalization factor. 
All the $\Fgc{g}$ transform as functions under symplectic
reparametrizations. Except for $g=1$, they   
are not holomorphic. The lack of holomorphy is governed 
by the following equation ($g>1$),
\beq
{\pa \Fgc{g} \over \pa \bar X^I}= {\textstyle{1\over 2}} \bar
F_I{}^{JK}  \sum_{r=1}^{g-1}\; {\pa \Fgc{r} \over \pa  X^J}\,{\pa \Fgc{g-r} 
\over \pa  X^K}\,, \label{anomalyeq2}
\eeq
where $\bar F_I{}^{JK}= \bar F_{ILM}\,N^{LJ}N^{MK}$. Observe that
this equation remains the same under $\Fgc{g}\to
\rho\,\m^{2g}\,\Fgc{g}$, corresponding to a rescaling 
\beq
F(X,W^2)\to \rho \,F(X,\mu^2\,W^2)\,, \label{rescaling}
\eeq
for arbitrary $\rho$ and $\mu$.

The integrability of (\ref{anomalyeq2}) imposes a condition on
$\Fg{1}$, 
\beq
\Big({\pa^2 \Fgc{1} \over \pa \bar X^I \pa X^K} \;  \bar
F_J{}^{KL} - (I\leftrightarrow J) \Big)   \,
{\pa \Fgc{g-1} \over \pa  X^L} =0\,. \label{aneqF1}
\eeq
This condition is trivially satisfied by (\ref{Fg}) as $\cFgc{1}$ 
defined by (\ref{Fg}) is holomorphic. An alternative 
solution of this equation is $\Fgc{1} \propto i [\bar
X^I\,F_I(X) - X^I\, \bar F_I(\bar X)]$. Note that the proportionality
constant in this solution can be adjusted by a rescaling of the
$\Fgc{g}$ of the type indicated above.

The above equations are applicable to both rigid and local 
supersymmetry. In the general case the coefficient functions are
just the derivatives of the full function $F$ with respect to a
chiral background field and the above results remain true even
without setting this
background equal to zero after applying the differentiations.
The latter is not so when we express the above results in terms
of the sections $X^I(z)$ defined in (\ref{section}), where
homogeneity of the coefficient functions 
is important, so that we are forced to set $W^2$ to zero. The
homogeneity of the $\Fg{g}$ in $X^I$ then ensures that the
$\Fgc{g}(X,\bar X)$ are  
homogeneous of degree $2(1-g)$ in $X$ and of degree 0 in $\bar
X$.  It is thus 
straightforward to construct sections $\cFgc{g}(z,\bar z)$
according to (\ref{Fsections}), which transform in the same way
as the $\cFg{g}$ 
under projective transformations (cf. \ref{Fprojective}). 
The lack of holomorphy is 
encoded in an equation which is rather similar to
(\ref{anomalyeq2}). It can be obtained by multiplying
(\ref{anomalyeq2}) by $\pa \bar X^I(\bar z)/ \pa \bar z^A$ and using
the identity
\beq 
N^{IJ} = e^{K} \Big[g^{A\bar B}\, (\partial_A+\partial_AK) X^I(z)\,
(\partial_{\bar B}+\partial_{\bar B}K) \bar X^J(\bar z) - X^I(z)\,\bar
X^J(\bar z) \Big]   \,,
\eeq
where $K(z,\bar z)$ and $g_{A\bar B}(z,\bar z)$ are the K\"ahler
potential and metric, respectively. The resulting equation, which 
we will refer to as the `symplectic' anomaly equation to distinguish it
from the anomaly equation discussed below, is covariant with 
respect to projective 
transformations and holomorphic diffeomorphisms and reads ($g>1$)
\beq
\pa_{\bar A} \cFgc{g}(z,\bar z) = {\textstyle{1\over 2}} e^{2K}
\bar {\cal W}_{\bar A}{}^{BC}\, \sum _{r=1}^{g-1} D_B
\cFgc{r}(z,\bar z)\,D_C \cFgc{g-r}(z,\bar z) \,. 
\label{anomalyeq3}  
\eeq
Here indices are raised or lowered by means of the K\"ahler metric
corresponding to (\ref{KP}). Covariant derivatives are
projectively covariant and defined by $D_A{\cal F}^{(g)} =
\big(\pa_A + 2 (1-g) \pa_A K\big) {\cal F}^{(g)}$; when acting on
tensors they include the Levi-Civita connection. Furthermore we
used the definition  
\beq
{\cal W}_{ABC}(z) =i  F_{IJK}\big(X(z)\big)  {\partial X^I(z)\over 
\partial z^A}
{\partial X^J(z)\over \partial z^B} 
{\partial X^K(z)\over \partial z^C} \,. 
\eeq
Although (\ref{anomalyeq3}) applies only to 
$g>1$, its integrability implies a condition for $\cFgc{1}(z,\bar
z)$ similar to (\ref{aneqF1}), with the $X^I$ replaced by 
$z^A$ and $\bar F_J^{\,KL}$ by  $\bar {\cal W}_{\bar 
A}{}^{BC}$.  

The above anomaly equations (\ref{anomalyeq3}) may be compared to the
anomaly equations derived some time ago in \cite{BCOV} for the
topological partition functions of twisted Calabi-Yau nonlinear
sigma models. They read,
\beq
\pa_{\bar A} \cFgc{g} = {\textstyle{1\over 2}} e^{2K} \bar {\cal W}_{\bar 
A}{}^{BC}\Big[\l^{-2}\, D_BD_C \cFgc{g-1} + \sum _{r=1}^{g-1} 
D_B \cFgc{r}\,D_C \cFgc{g-r}\Big] \,,  \label{anomalyeq} 
\eeq
for $g>1$, whereas for $g=1$  we have
\beqa
\pa_A\pa _{\bar B} \cFgc{1} &=&\l^{-2}\,\Big[ {\textstyle{1\over 2}} e^{2K}\,
{\cal W}_{ACD}  
\bar{\cal W}_{\bar B}{}^{CD} +\big(1- {\textstyle{1\over
    24}}\chi \big) \, g_{A\bar B}  \Big]
\nonumber \\ 
&=&\l^{-2}\,\Big[- {\textstyle{1\over 2}} R_{A\bar B} - {\textstyle{1\over
24}}\big[\chi -12(n+3)\big] g_{A\bar B}\Big] \;, 
\label{nonholoF1}
\eeqa
where $R_{A\bar B}$ denotes the Ricci tensor of the Calabi--Yau moduli space
and  $\chi$ the Euler number  
of the Calabi-Yau.\footnote{%
   A particular solution of  
   (\ref{nonholoF1}) is $\cFgc{1}= \l^{-2}\,\big[ -{\textstyle{1\over 2}} \ln
   g -{\textstyle{1\over 24}}\big[\chi -12(n+3)\big]\big] K$, where $g$ is the
   determinant of the K\"ahler metric. } %
The value of the  coefficient $\l^2$ depends on the normalization
adopted for the $\cFgc{g}$, as follows from performing the rescaling
(\ref{rescaling}). The constants
$\l^{2g}$ measure the 
strength of the genus-$g$ partition function so that, in the
context of type-II string theory, $\l^2$ can be identified with 
the type-II string--coupling constant. 

Clearly (\ref{anomalyeq2}) may be regarded as a truncation (for 
instance, arising from the singular limit $\l\to \infty$) of
(\ref{anomalyeq}). The sections
constructed from (\ref{Fg}) are a solution of the truncated anomaly 
equation (\ref{anomalyeq3}) and are unique provided that
$\cFgc{1}$ is taken to be holomorphic. Non-holomorphic corrections to
$\cFgc{1}$ can 
be included separately and they will propagate into the
higher-$g$ coefficient functions upon solving the appropriate
anomaly equation (i.e., (\ref{anomalyeq3}) or
(\ref{anomalyeq})). This lack of uniqueness does not represent a
problem of principle. The requirement
of constituting a function under symplectic reparametrizations
cannot uniquely determine the non-holomorphic terms and the
construction based on (\ref{Fg}) generates the non-holomorphic
modifications that are 
minimally required in order to extend the Wilsonian coefficient
functions into symplectic functions.

In a given holomorphic parametrization of the sections $X^I(z)$ 
the symplectic transformations induce a corresponding 
transformation on the coordinates $z$. In order to remain within 
a given gauge, this transformation is accompanied by a projective 
transformation,  
\beq
X^I(z) \longrightarrow \tilde X^I(\tilde z)= e^{f(z)}\, 
\Big[U^I_{\,J} \, X^J(z) +Z^{IJ} F_J\big(X(z)\big) \Big] \,.  
\label{ztrans} 
\eeq
With these definitions we obtain the following 
transformation rule for the $\cFgc{g}$, 
\beq
\tilde\cFgc{g}(\tilde z,\bar {\tilde z})=  
e^{2(1-g)f(z)}\,\cFgc{g}(z,\bar z)\,. \label{Ftrans}
\eeq

As mentioned before, the anomaly equation (\ref{anomalyeq}) was
derived for the genus-$g$ partition functions of twisted
Calabi--Yau sigma models. They were shown to
correspond to certain type-II $g$-loop string amplitudes in
\cite{AGNT1}. It is worthwhile to indicate the origin of the
various terms in the anomaly equation. Generically the
defining expressions for the $\cFgc{g}$ are holomorphic. However,
when integrating over the moduli space of genus-$g$ Riemann
surfaces one encounters boundary terms associated with various pinchings
of the Riemann surface. The first term in (\ref{anomalyeq}) is
due to the pinching of one of the homology cycles, which explains
why the genus is lowered by one unit; the second term correspond
to a pinching that leads to two disconnected surfaces, so that
the sum of their genera equals the original genus $g$. In terms of
the effective field theory, these pinchings are identified as the
effects of the propagation of massless modes \cite{DKL,AGNT1}. As is
well known, these effects form an obstacle in obtaining a local
effective action. The Wilsonian action, on the 
other hand, is a local
effective action, in which the cumbersome effects of the massless
modes are avoided by the presence of an infrared
cut--off. However, because of this cut--off the Wilsonian action
does not fully capture the physics, and certain features (like
the presence of certain symmetries) of the underlying model are
not always manifest.

The approach based on (\ref{Fg}) encapsulates some of these
features. The Wilsonian coefficient functions are not covariant
with respect to the symplectic reparametrizations and therefore
will not be  
invariant under certain subgroups (such as target-space
duality), as one would expect from an {\it ab initio}
calculation based on an underlying physical theory that has this
invariance. Certain nonholomorphic corrections readjust this 
situation, but they themselves have no role to play in the
Wilsonian set--up. Identifying these nonholomorphic corrections
with the contributions from propagating massless modes provides an
explanation for this phenomenon. Comparison with the anomaly
equation (\ref{anomalyeq}) of \cite{BCOV} indicates that the
approach based on 
(\ref{Fg}) correctly takes into acount the massless--tree
contributions. The massless loops, while not excluded in this
approach, will appear as separate contributions. 
As we will discuss in the next section, in the semi-classical limit
of the heterotic string only the second term in the anomaly equation
(\ref{anomalyeq}) survives; hence for the perturbative heterotic string
the lack of holomorphy is fully governed  by
(\ref{anomalyeq3}). Interestingly enough, as alluded to earlier,
the same effect takes place 
on the type-II side in the strong-coupling limit. The
expressions (\ref{Fg}) represent   
an explicit solution to the latter anomaly equation.

To elucidate the construction based on (\ref{Fg}), we derive the
first few covariant functions $\cFgc{g}$ for a specific example,
where the $W^2$-independent part of the holomorphic function
equals  
\beq
F(X,W^2)\big|_{W^2=0}= {d_{ABC}\,X^AX^BX^C\over X^0}\,. \label{dfunction}
\eeq
Its corresponding K\"ahler potential takes the form 
\begin{equation}
K(z,\bar z)=-\log \Big(-id_{ABC}\,(z-\bar
z)^A(z-\bar z)^B(z-\bar z)^C\Big)  \,. \label{dKpotential}
\end{equation}
Here we employ so-called special coordinates $z^A$ by choosing
the holomorphic sections $(X^0(z), X^A(z)) = (1,z^A)$. The matrix
$N_{IJ}\equiv -i (F_{IJ}  -\bar F_{IJ})$ is then equal to  
($I= 0,A$)
\beq
N_{IJ}= \pmatrix{2n_{CD}(z^Cz^D+z^C\bar z^D +\bar z^C\bar z^D) & 
-3n_{BC}(z+\bar z)^C\cr \noalign{\vskip3mm}
-3n_{AD}(z+\bar z)^D& 6 n_{AB}\cr} \,,
\eeq
where $n_{AB}= -id_{ABC}\,(z-\bar z)^C$, and its inverse equals  
\beq
N^{IJ} = \pmatrix{2\,e^K  & 
e^K\,(z+\bar z)^B \cr\noalign{\vskip 3mm}
e^K\, (z+\bar z)^A  &{1\over6} n^{AB} +{1\over 2}e^K\,(z+\bar 
z)^A(z+\bar z)^B \cr} \,. \label{inverseN}
\eeq

So far we put $W^2=0$. We now consider the $W^2$-dependent terms
and construct the covariant coefficient functions by using
(\ref{Fg}). With the exception of the first one, 
\beq
\cFgc{1}(z,\bar z) = \cFg{1}(z)\,,
\eeq
all other functions are nonholomorphic. We exhibit the explicit
expressions for $\cFgc{2}(z,\bar z)$ and $\cFgc{3}(z,\bar z)$, 
\beqa
\cFgc{2}(z,\bar z) &=& \cFg{2} (z) + \textstyle{1\over12} 
\wh{n}^{AB}
\pa_A\cFg{1}(z)\,
\pa_B\cFg{1}(z) \,,\nonumber \\
\cFgc{3}(z,\bar z) &=& \cFg{3}(z) 
+ \textstyle{1\over6} \wh{n}^{AB} \; \pa_A\cFg{2}(z)\,
\pa_B\cFg{1}(z)   \nonumber\\
&& + 2 e^K\,(z-\bar z)^A\pa_A \cFg{1}(z) \; \cFg{2}(z)  \nonumber\\
&& +{\textstyle{1\over72}} \wh{n}^{AC} \,\wh{n}^{BD}\,
\pa_A\pa_B\cFg{1}(z)\;\pa_C\cFg{1}(z)\;\pa_D\cFg{1}(z)\nonumber\\
&&+ {\textstyle{1\over6}} e^K\, \wh{n}^{AB} \, (z-\bar z)^C 
\pa_A\cFg{1}(z)\; \pa_B\cFg{1}(z) \; \pa_C \cFg{1}(z) 
\nonumber\\ 
&&
+{\textstyle{i\over216}}  d_{ABC}\, \wh{n}^{AD} \, \wh{n}^{BE} \,
\wh{n}^{CF}\, \pa_D\cFg{1}(z)\;\pa_E\cFg{1}(z)\;\pa_F\cFg{1}(z)\,,  
\label{thirdfc}
\eeqa 
where $\wh{n}^{AB} = n^{AB} + 3 \,e^K \, (z- \ov{z})^A \, (z -
\ov{z})^B$. 
Not surprisingly, the resulting expressions are rather similar
to the ones in the orbifold example in section~7.1 of
\cite{BCOV}, which are, however, based on a non-holomorphic
$\cFgc{1}$ and define the solutions of  
the anomaly equation (\ref{anomalyeq}) subject to modular invariance. 

Unlike the holomorphic quantities $\cFg{g}$, these non-holomorphic 
quantities $\cFgc{g}$ transform as  
functions under symplectic reparametrizations. They satisfy the 
holomorphic anomaly equation (\ref{anomalyeq3}). In
appendix~\ref{fgstu} we present the above results for the case of
the $S$-$T$-$U$ model.


\section{The holomorphic anomaly 
equations and their solutions for the heterotic string}
\setcounter{equation}{0}

In this section we focus on $N=2$ supersymmetric models in four
space--time dimensions that have both a type-II and
a heterotic description \cite{KV}.  In the type-II description,
such models are obtained by compactifications of the type-IIA string on
certain Calabi--Yau manifolds.  In the heterotic description,
they follow from 
compactifications of the heterotic $E_8 \times E_8$ 
string on $K_3 \times T_2$. String--string duality
then implies that 
the non-holomorphic couplings $\cFgc{g}(z,{\bar z})$ in the
type-II and in the heterotic description are related. On the
type-IIA side, the $z^A$ denote the K\"ahler-class 
moduli, whereas on the heterotic side the $z^A$ correspond to the
heterotic dilaton $S$ and to the heterotic moduli $T^a$  
(consisting of the toroidal and Wilson-line moduli). Partial
evidence for such a string--string 
duality has been given for the case of 2 moduli in
\cite{KLT,AGNT2,Curio}, where on the heterotic side we have the
complex dilaton field $S$ and a modulus $T$, and for the case of
3 moduli in \cite{KLT,Curio,CCLM}, with the dilaton $S$
and the two $T_2$ moduli $T$ and $U$ on the heterotic side. This
model, which  we will refer to 
as the $S$-$T$-$U$ model, will be discussed in more detail
later in this section. More recently, evidence for string--string
duality was also obtained for
the case of 4 moduli, which on the heterotic side incorporates the two
toroidal moduli $T$ and $U$ and a Wilson-line modulus \cite{BKKM}.  

In the context of type-II compactifications on
Calabi--Yau manifolds, it was shown in \cite{AGNT1}
that the non-holomorphic 
sections ${\cal F}^{(g){\rm cov}}(z ,\bar{z})$
obtained from direct string calculations are equal to the
topological partition functions, and thus they satisfy 
the holomorphic anomaly equations (\ref{anomalyeq}) and
(\ref{nonholoF1}). In type-II string compactifications the  
contributions to the $\cFgc{g}$ originate from $g$ loops 
in string perturbation theory \cite{AGNT1}. This can be seen as follows. 
According to (\ref{Fsections}), the $\cFgc{g}$ are multiplied by a
factor $(\mpl^2)^{1-g}$; 
keeping the string scale rather than the Planck scale fixed  
yields a factor $[g_{\rm s}^{-2}]^{g-1}$, where $g_{\rm 
s}^{-2}$ is proportional to the dilaton and acts as a 
loop-counting parameter. There can be no further dependence on
the string coupling, as the type-II dilaton resides in a
hypermultiplet and neutral hypermultiplets do not affect the
vector-multiplet couplings. Thus we are dealing with
contributions at precisely $g$ loops. 
Note that the relevant anomaly equation (\ref{anomalyeq}) 
indeed comprises terms of the same loop order. The first term 
describes the $(g-1)$-loop contribution with a   
massless loop appended to it, while the second term describes the 
product of an $r$--oop and a $(g-r)$-loop contribution.

For type-II models possessing a dual heterotic description,
one thus expects that the heterotic couplings
satisfy similar holomorphic anomaly equations. 
As the arguments in the previous
section have shown, the existence of such anomaly equations can,
at least partially, be understood from arguments based on symplectic
reparametrizations, from which one may conclude that certain
features concerning the 
non--holomorphic terms should be generic and independent of the
precise model one is considering. This observation will 
help us to fix the relative normalization between the sections
obtained  on the heterotic and on the type-II side. In fact,
as we will demonstrate shortly, in the relevant limit of a large 
K\"ahler-class modulus the type-II and the symplectic anomaly equations 
become identical.

Later in this section we will turn to the heterotic weak-coupling
limit of these holomorphic 
anomaly equations and we will solve them for $\cFgc{2}$ and 
$\cFgc{3}$ in the context of the $S$-$T$-$U$ model, for
concreteness.  The $\cFgc{g}$ exhibit singularities
at lines/points in the perturbative heterotic moduli space
at which one has perturbative gauge--symmetry enhancement.  
We will show that, in the vicinity of these lines/points
of semi-classical gauge symmetry enhancement, the structure of 
the $\cFgc{2}$ and $\cFgc{3}$ we obtain 
precisely agrees
with expressions (\ref{thirdfc}) (with certain one-loop
corrections included) found in the previous section on the
grounds of symplectic covariance. Subsequently we analyze the 
target--space duality properties and exploit the covariance to further 
restrict the couplings in terms of modular forms.


\subsection{The weak-coupling limit in the heterotic string and
  target--space duality}

As discussed above, in type-II string compactifications the 
contributions to the $\cFgc{g}$ originate from $g$ loops 
in string perturbation theory. For heterotic vacua the counting is 
different, because the dilaton resides in the vector multiplet 
sector. First of all, the K\"ahler potential contains a 
characteristic term $\ln g^2_{\rm s}$, so that the factor
$(\mpl^2)^{g-1}\exp [-(1-g)K]$ in the definition
(\ref{Fsections}) of the holomorphic sections $\cFg{g}$ depends
only on the string scale and no longer on $g_{\rm s}$. Therefore
the dependence of the corresponding couplings on $g_{\rm s}$ is
directly related to the explicit 
dependence of the $\cFg{g}$ on the dilaton. The presence of the
dilaton is, however, restricted by 
non-renormalization arguments. As a result, only 
the first two terms in the expansion (\ref{holoexpansion}) depend 
explicitly and linearly on the dilaton field (in perturbation 
theory), and thus represent tree-level contributions. All 
other terms contribute at precisely one loop in string 
perturbation theory \cite{AGNT1}. 

Let us now assume that the heterotic sections $\cFgc{g}$ satisfy
a holomorphic anomaly equation similar to (\ref{anomalyeq}) and
deduce, on the basis of the counting arguments given
above, what the relevant terms will be in the weak-coupling limit
$\modS\to\infty$. Since generically all the 
$\cFg{g}$ are independent of $S$ and thus correspond to one-loop
contributions, it follows  that the  
right-hand side of the anomaly equation is generically of two-loop 
order and will therefore be suppressed by a factor  
$g_{\rm s}^2$: the first term in the anomaly equation appends a 
massless loop to a one--loop term, while the second term consists 
of products $\cFg{g-r}\, \cFg{r}$ of one-loop
terms. Nevertheless, one-loop
contributions  can still emerge from $\cFgc{1}$, which is not  
exclusively the result of a one-loop correction but  
contains also a term arising from the tree
approximation. However, this term is linear in the dilaton, so
that the second derivative term in (\ref{anomalyeq}) cancels and
we are left with the truncated equation
(\ref{anomalyeq3}). Actually, also
this term simplifies, as it generically contributes at the
two--loop level, with the exception of the terms proportional to
$\partial_S\cFgc{1}\,\partial_a\cFgc{g-1}$ 
and, for $g=2$, $\partial_S\cFgc{1}\,\partial_S\cFgc{1}$, which
can still give rise to one-loop contributions. 

To confirm the above argument let us explicitly consider the
limit of large $\modS$ in the anomaly equation. We consider the
dilaton $S= 4 \pi /g^2 -i \theta/ 2 \pi$ 
and an arbitrary number of moduli $T^a$, which are
related to the special coordinates $z^A$ by $iS= z^1$, $i
T^a=z^a$. The class of compactifications is defined by the 
requirement that, up to nonperturbative contributions which take
the form of positive powers of $\exp(-2\pi S)$, the 
associated holomorphic prepotentials are given by 
\beqa
{\cal F}^{(0)}(S,T^a) = - S \,T^a \eta_{ab} T^b + h(T^a) \,,
\label{cFnull}
\eeqa
and the Wilsonian couplings ${\cal F}^{(g)}$ are given by
\beqa
{\cal F}^{(1)}(S,T^a)) &=&  a \,S  + h^{(1)}(T^a) \;, \nonumber\\
{\cal F}^{(g>1)}(S,T^a) &=& {\cal F}^{(g)}(T^a) \;, 
\label{fg2}
\eeqa
where $a$ denotes an integer.\footnote{We will, throughout the paper,
  use the normalization convention $a=24$ \cite{AL}, thereby fixing the
  normalization of $W^2$.
  \label{normalization}}
The K\"ahler potential
for the above models, which is computed from (\ref{cFnull}), is given by
\beq
K(S,T) = - \log(S + \bar{S} + V(T,{\bar T} )) + \hat K(T,\bar
T)\,, \label{Kpotential}
\eeq
where the K\"ahler potential $\hat K$ and the corresponding
K\"ahler metric are given by  
\beqa  
\hat K(T,\bar T) &=& -\log[ (T+\bar T)^a\eta_{ab} (T+\bar
T)^b] \,,\nonumber \\
\hat g_{a\bar b} &=& -2 \eta_{ab}\,e^{\hat K} + 4
\eta_{ac}\eta_{bd} (T+\bar T)^c  (T+\bar T)^d\, e^{2\hat K} \,, \nonumber \\
\hat g^{a\bar b} &=& -{\textstyle{1\over 2}}\eta^{ab} \,e^{-\hat
  K} + (T+\bar T)^a  (T+\bar T)^b  \,, \label{hatkahler}
\eeqa
where $\eta^{ab}$ is the inverse of $\eta_{ab}$. The quantity $V$
is the Green-Schwarz term, defined by 
\beq
V(T,\bar T)= {2(h+\bar h) -(T+\bar T)^a (h_a+\bar h_{\bar
    a})\over (T+\bar T)^a \eta_{ab}(T+\bar T)^b}\,. \label{defGS}
\eeq
This term satisfies the following equation, which will be useful later on, 
\beqa
&&\Big[\pa_a\pa_{\bar b} + \pa_b\pa_{\bar a}+4
\eta_{ab}\,e^{\hat K} - 16 \big[\eta(T+\bar T)\big]_a
\big[\eta(T+\bar T)\big]_b\,e^{2\hat K}\Big] V = \nonumber \\
&& - 2(h_{ab} +\bar h_{\bar a\bar b}) e^{\hat K}  
+2\Big[ \big[\eta(T+\bar T)\big]_a (h_{bc} +\bar h_{\bar b\bar
c})+ (a\leftrightarrow b) \Big] \,(T^c+\bar T^c)\, e^{2\hat K} \,, 
\label{GSidentity}
\eeqa
where $\big[\eta(T+\bar T)\big]_a=\eta_{ab}\,(T+\bar T)^b$.

The behaviour of
the Wilsonian couplings ${\cal F}^{(g)}$ in the limit $\modS\to
\infty$ can easily be determined from (\ref{fg2}). 
However, the functions entering
in the holomorphic 
anomaly equation (\ref{anomalyeq}) are not the Wilsonian coupling
functions, but rather the 
full non-holomorphic functions ${\cal F}^{(g) {\rm cov}}(z,{\bar z})$.
Nevertheless, let us momentarily assume that they satisfy the
following conditions in the limit $\modS
\rightarrow  \infty$, which are somewhat weaker but consistent
with what one would derive for the Wilsonian couplings on the
basis of (\ref{fg2}), 
\beqa
D_S {\cal F}^{(g>1)\,{\rm cov}} 
&\rightarrow& 0 \;, \nonumber\\
D_S {\cal F}^{(1)\,{\rm cov}} &\rightarrow& a \;,\nonumber\\
D_{\bar S} {\cal F}^{(g\geq1)\,{\rm cov}} 
&\rightarrow& 0 \;, \nonumber\\
D_{T} {\cal F}^{(g\geq1)\,{\rm cov}} 
&\rightarrow& f^{(g)} (T,{\bar T}) \;,
\label{assum}
\eeqa
where $f^{(g)}$ is some arbitrary function. 
Let us now consider
the anomaly equation 
(\ref{anomalyeq}) in the limit $\modS  \rightarrow
 \infty$.  Because the only non-zero
components of the tensor ${\cal W}_{ABC}$ are ${\cal W}_{abS}$
and ${\cal W}_{abc}$, the anomaly equation takes the form
\beqa
\pa_{\bar a} \cFgc{g} &=& e^{2K} \bar {\cal W}_{\bar 
a\bar b \bar s}\Big[ D^{\bar b}D^{\bar s} \cFgc{g-1} +
\sum_{r=1}^{g-1}   
D^{\bar b} \cFgc{r}\,D^{\bar s} \cFgc{g-r}\Big] \nonumber \\
&&+{\textstyle{1\over 2}} e^{2K} \bar {\cal W}_{\bar 
a\bar b \bar c}\Big[ D^{\bar b}D^{\bar c} \cFgc{g-1} +
\sum_{r=1}^{g-1}D^{\bar b} \cFgc{r}\,D^{\bar c} \cFgc{g-r}\Big]
\,.  
\eeqa   
Using the results (\ref{DFDF}) and (\ref{DDF}) of appendix
\ref{sbig} and the asymptotic conditions (\ref{assum}), it is then 
straightforward to show that, in the limit $\modS \rightarrow \infty$,
the $D_BD_C \cFgc{g-1}$ term does not contribute anything in the
anomaly equation (\ref{anomalyeq}), which is therefore reduced to
the equation (\ref{anomalyeq3}). Actually, this equation reduces
to an even simpler form ($g>1$), 
\beq
\pa_{\bar a} \cFgc{g} = e^{2\hat K} \bar {\cal W}_{\bar 
a\bar b \bar S}\,\hat g^{\bar b c}\Big[a \, \big(
\pa_{c} +
2(1-(g-1))\pa_c\hat K\big) \cFgc{g-1} - 
a^2\, V_c\, \d_{g,2} \Big] \,,  \label{anomsbig} 
\eeq
where $\partial_a = \pa/\partial{T^a}$. As exhibited in
(\ref{metricsbig}), the quantity $-\hat g^{\bar a b} \,V_b$, which appears 
only at genus-2, is equal to the
inverse metric component $g^{\bar a S}$ in the limit of large
$\modS$. 
The consistency of the assumption made above that the $\cFgc{g}$ exhibit a
similar behaviour at large $\modS$ as the holomorphic sections
$\cFg{g}$, is confirmed by considering the holomorphic anomaly
equation for $\partial_{\bar S} \cFgc{g}$, whose right-hand side
behaves as $(\modS)^{-2}$ in the large $\modS$ limit, again as a
result of (\ref{DFDF}), (\ref{DDF}) and
(\ref{assum}).
For the case of two moduli (i.e. only $S$ and $T$) the above
result (\ref{anomsbig}) was already derived in \cite{AGNT2}.

Of course, when the models that we are considering have both a heterotic 
and a type-II description, the covariant sections 
${\cal F}^{(g){\rm cov}}$ are  
identical to each other, upon a suitable identification
of the type-II and the heterotic fields.  
A priori it may not be clear that the normalization of the
covariant sections is the same. However, in the semi-classical
limit we have established that they are subject to the same 
anomaly equation (\ref{anomalyeq3}) which, according to (\ref{rescaling}),  
allows for different
normalizations corresponding to only an overall rescaling of the full
Wilsonian function $F(X, W^2)$, and of the field $W$. The first
one is fixed once the $W^2$-independent part of $F$ is fixed. 
The second one simply depends on the normalization adopted for
$W$ and is thus related to the relative normalization of the
topological partition functions (i.e., the parameter $\l$ in
(\ref{anomalyeq})) vis \`a vis the normalization of the Weyl
multiplet in the Lagrangian for the effective field theory on the
heterotic side. On the type-II side this connection is
provided by the work of \cite{AGNT1}, which showed  that certain 
type-II string amplitudes precisely reproduce the topological partition
functions. Our analysis is entirely on the heterotic side and we
will simply adopt the standard normalization convention for the Weyl
multiplet leading to $a=24$ (cf. footnote \ref{normalization}).

Let us consider the anomaly equation for $\cFgc{2}(T,{\bar T})$
in more detail.  At one-loop, $\cFgc{1}$ will be given by
\beqa
\cFgc{1} = a S_{\rm inv} + h^{(1)\,{\rm cov}}_{\rm inv}(T,{\bar
  T})\; , \label{tildef1}
\eeqa
where $S_{\rm inv}$ denotes the so-called invariant dilaton
\cite{DWKLL}. It differs from $S$ by a holomorphic function
$\sigma(T)$ of the $T^a$,
\beq
S_{\rm inv}-S\equiv \sigma(T) =  \frac{1}{2(n+1)}\Big(- \eta^{ab}
h_{ab}(T)  +L(T) \Big)\,.
\eeq
Here $L$ is a holomorphic 
function of the moduli fields $T^a$ which transforms into
imaginary constant shifts under target--space duality
transformations. These shifts are associated with semiclassical
monodromies. The new dilaton field $S_{\rm inv}$ is 
invariant under target--space duality transformations, but it is
no longer a {\it special} coordinate in the context of $N=2$
supersymmetry. The true loop-counting parameter 
of the heterotic string, on the other hand, is given by \cite{DWKLL}
\beqa
S + {\bar S} + V(T,{\bar T}) = S_{\rm inv} + {\bar S}_{\rm inv} + 
V_{\rm inv}(T,{\bar T})
\label{loopc}
\eeqa
where $V_{\rm inv}(T,{\bar T})$ denotes the so-called
invariant Green--Schwarz term, defined by this equation.  
$V_{\rm inv}(T,{\bar T})$ is invariant under one--loop target--space
duality transformations.  It follows from (\ref{loopc}) that 
$V(T,{\bar T}) = V_{\rm inv}(T,{\bar T}) + \sigma (T) + 
{\bar \sigma} ({\bar T})$.  Hence,
\beqa
\partial_b \Big[
\cFgc{1} - a V(T,{\bar T})\Big] = \partial_b \Big[
h_{\rm inv}^{(1)\,{\rm cov}}(T,{\bar T}) - a V_{\rm inv} (T,{\bar T})
\Big] \;\;.
\eeqa
The holomorphic anomaly equation for $\cFgc{2}(T,{\bar T})$
can then be rewritten into
\beqa
\pa_{\bar a} \cFgc{2} &=& a \,e^{2\hat K} \bar {\cal W}_{\bar 
a\bar b \bar S}\, \hat g^{\bar b c} \,\pa_{c} \Big[ \cFgc{1} - 
a \, V  \Big] \nonumber\\
&=& 
a \,e^{2\hat K} \bar {\cal W}_{\bar 
a\bar b \bar S} \,\hat g^{\bar b c} \,\pa_{c} \Big[ h_{\rm
inv}^{(1)\, {\rm cov}} - a \, V_{\rm inv}  \Big] \,.
\label{holdf2}
\eeqa
As the metric and the tensors $\cal W$ must be target-space
duality covariant, this exhibits the manifest covariance of the
anomaly equation in the perturbative limit. The correctness of this 
result can be inferred from the covariance of the original 
anomaly equation in the context of the $(\modS)\to\infty$ limit.


\subsection{The anomaly equation for the heterotic $S$-$T$-$U$ model}

The $S$-$T$-$U$ model can be constructed by compactifying
the ten-dimensional heterotic string on $T_2\times K_3$.
A compactification of the $E_8\times E_8$ heterotic string
on $K_3$, with equal $SU(2)$
instanton number in both $E_8$ factors, gives
rise to a model in six space--time dimensions with gauge group
$E_7\times E_7$. For general vacuum expectation values
of the massless hypermultiplets this gauge group is completely
broken, and one is left with 244 massless hypermultiplets and no massless
vector multiplets. Upon a $T_2$ compactification down to four dimensions,
one obtains a model with 244 massless hypermultiplets and with
three massless vector multiplets $S$, $T$ and $U$, 
where $S$ denotes the heterotic dilaton and $T,U$ denote the
$T_2$ moduli. This model is the heterotic dual of the type IIA--model 
\cite{KV} based on the  Calabi-Yau space 
$WP_{1,1,2,8,12}(24)$ with $h_{11}=3$, $h_{21}=243$ and, hence, 
with $\chi=-480$.

In the heterotic description this model possesses a target--space
duality symmetry $SL(2,{\bf Z})_T\times 
SL(2,{\bf Z})_U\times {\bf Z}_2^{T\leftrightarrow U}$ at the
perturbative level ($\modS\rightarrow\infty$). All elements of
this discrete symmetry group act as symplectic transformations of the form
(\ref{Xtrans}) subject to condition (\ref{invariantF}). 
Hence, according to (\ref{Ftrans}) the non-holomorphic functions
$\cFgc{g}$  transform as modular functions of a specific modular
weight.  As it is well known, the duality transformations
$T\rightarrow \tilde T={aT-ib\over icT+d}$ 
induce a particular K\"ahler transformation of the K\"ahler potential, 
$K \rightarrow K - f - \bar{f}$, with $f=-\log (icT+d)$.
Then, comparison with (\ref{Ftrans}) shows that the
non-holomorphic $\cFgc{g}$ possess the following duality
transformation behaviour,
\beqa
\cFgc{g}(\tilde T,\bar {\tilde T},U,\bar U) \rightarrow (icT+d)^{2(g-1)}
\cFgc{g}(T,\bar T,U,\bar U)\,,
\label{mwstu}
\eeqa
and similarly for $U \rightarrow  \tilde U = \frac{aU-ib}{icU+d}$.
Note that, according to
(\ref{tildef1}), $h^{(1){\rm cov}}_{\rm inv}(T,{\bar T}, U,
{\bar U})$ and $\cFgc{1}(T,{\bar T}, U,
{\bar U})$ are both invariant 
under target--space duality transformations (\ref{mwstu}).

The perturbative prepotential for the $S$-$T$-$U$ model reads 
\beqa
{\cal F}^{(0)}(S,T,U) = -STU + h(T,U) \;\;,\label{hetprep}
\eeqa
where $h(T,U)$ denotes the one-loop contribution to the prepotential; 
it would transform as a modular form of weight $-2$, were it not for 
the presence of an inhomogeneous term in its transformation rule 
proportional to a polynomial of $T$ and $U$ containing no powers 
higher than $T^2$ or $U^2$. This polynomial is directly related 
to the semi-classical monodromies. It then follows \cite{DWKLL} that 
$\pa_T^3 h$ and $\pa_U^3 h$ are single-valued modular functions of 
weight $+4$ and $-2$, respectively, under $SL(2,{\bf Z)}_T$, and of 
weight $-2$ and $+4$, respectively, under $SL(2,{\bf Z})_U$. 

Using the results (\ref{metricsbig}) given in appendix \ref{sbig} and the
explicit form for the matrix $\eta_{ab}$,  one finds
that, in the limit $\modS \rightarrow \infty$,
\beqa
g^{S\bar S}=(S + \bar S)^2 \;\;,\;\; g^{T \bar T}=(T + \bar T)^2 \;\;,\;\;
g^{U \bar U}=(U + \bar U)^2 \;\;, \nonumber\\
g^{T \bar S}= - (T + \bar T)^2 \,\partial_T V(T,\bar T,U,\bar U) \;\;,\;\;
g^{U \bar S}=  - (U + \bar U)^2 \,\partial_U V(T,\bar T,U,\bar U)\;, 
\label{metric}
\eeqa
with the Green--Schwarz term $V(T,\bar T,U,\bar U)$ given as 
\beqa
V(T,\bar T,U,\bar U)\ =\ {2(h+\bar h)\,
-\,(T+\bar T)(\partial_T h+\partial_{\bar T}\bar h)\,
-\,(U+\bar U)(\partial_U h+\partial_{\bar U}\bar h) 
\over (T+\bar T)\,(U+\bar U)} \;,
\label{GSfunction}
\eeqa
as well as ${{\cal W}}_{{ S}{T}{ U}} =-1$.
Then, it follows from (\ref{anomsbig}) and from (\ref{holdf2})
that 
\beqa 
{\partial}_{\bar T} \cFgc{g} &=&  -\frac{a}{(T + \bar T)^2} 
D_U \cFgc{g-1}\,,\quad g>2\,,\label{fghetweak} \\
{\partial}_{\bar T} \cFgc{2}  &=&
-\frac{a}{(T + \bar T)^2} 
\partial_U \left(h^{(1){\rm cov}}_{\rm inv}- a\,V_{\rm inv}
\right)  \,, \label{f2het}
\eeqa
and likewise with $T$ and $U$ interchanged.
Note that both $h^{(1){\rm cov}}_{\rm inv}$ and $V_{\rm inv}$
are target--space duality invariant, so that both
(\ref{fghetweak}) and (\ref{f2het}) are consistent with the
modular transformation behaviour given in (\ref{mwstu}).

At $T = U$ one has a perturbative gauge--symmetry enhancement,
so one expects that 
the $\cFgc{g}$ are singular as $T \rightarrow U$.  The leading singularities
will be holomorphic, whereas the non-leading singularities
will be both of the holomorphic and of the non-holomorphic type.
In the following subsection we turn to the discussion of these 
leading holomorphic singularities.


\subsection{Leading holomorphic singularities in heterotic models}

The holomorphic Wilsonian couplings ${\cal F}^{(g)}(z)$ are
singular at precisely those points in the moduli space where certain
string states become massless. In the type-II context,
for example, precisely
one hypermultiplet becomes massless at the conifold points
of the Calabi-Yau moduli space \cite{Strom},
and the leading holomorphic singularity follows from a $c=1$
matrix model and is given by \cite{GosVa} 
\beqa
{\cal F}^{(g>1)}(z)
=  \frac{B_{2g}}{2g(2g-2)}\frac{1}{[{\cal \mu}(z)]^{2(g-1)}}\;\; , 
\label{sinfgh}
\eeqa
where $\mu(z)$ is the mass of the hypermultiplet.  The $B_{2g}$
are the Bernoulli numbers.

Let us now discuss the leading singularity structure of the
perturbative Wilsonian couplings 
${\cal F}^{(g)}(z)$ in the heterotic context by
taking into account the moduli-dependent, elementary string spectrum.
We base ourselves on the perturbative prepotential
(\ref{cFnull}). 
The mass formula for a BPS state equals  
$m^2_{BPS} \propto  \mpl^2\, e^{K(z,\bar z)} |{\cal M}(z)|^2$, where 
${\cal M}(z)$ is a holomorphic section defined by
\beq
{\cal M}(z) \equiv M_I X^I(z) -  N^I F_I(z)\,,
\eeq
and  $M_I$ and $N^I$ denote the electric and the magnetic charges
of a given BPS state. Obviously, ${\cal M}(z)$ transforms under projective
transformations as ${\cal M}(z)\to e^{f(z)}\,{\cal M}(z)$. The
separation into `electric' and 
`magnetic' charges refers to a specific symplectic basis. In the
basis that is relevant for classical string theory
\cite{Sen,CAFP,DWKLL,AFGNT}, the symplectic holomorphic sections
read $(\hat X^I(z),\hat F_I(z))= ( 1, \eta_{bc}T^bT^c, iT^a
,-iS\,\eta_{bc}T^bT^c-2ih +iT^b \,\pa_b h,-iS,2S\,\eta_{ab}T^b
-\pa_a h)$. With this result the perturbative holomorphic mass  
${\cal M}(S,T^a)$ equals  
\beqa
{\cal M}(S,T)  &=& M_0 +  M_1\, \eta_{ab}T^aT^b + i M_a T^a  \\
&&+i S (N^0 \,\eta_{ab}T^aT^b   + N^1 + 2i \eta_{ab} N^a \,T^b)
+2iN^0\,h -(iN^0 T^a -N^a) \,\pa_ah \;. \nonumber
\eeqa
For elementary string states (i.e., states whose mass remains finite in the
classical limit), $N^I = 0$.  
Hence, 
the holomorphic mass ${\cal M}(T)$ for elementary string states
is given by 
\beqa
{\cal M}(T) = M_0 + M_1 \,\eta_{ab}T^aT^b  + i M_a\, T^a \,,
\eeqa
which is not affected by perturbative corrections.

Now, let us briefly consider a simple heterotic model
with just one modulus field $T$, the so-called $S$-$T$ model.
In this model, the holomorphic mass for elementary string states
can vanish at $T=1$.  At this point in the perturbative
moduli space, the $U(1)$ associated with the $T$ modulus becomes
enhanced to an $SU(2)$, and thus
two additional vector multiplets become massless.
In the vicinity of $T=1$, the $\cFg{g}$ should have
a leading holomorphic 
singularity given by
\beqa
\cFg{g>1}(T)
\propto  \frac{1}{{\cal M}^{2(g-1)}(T)}  \;\;, \label{masssin} 
\eeqa
where ${\cal M} \propto T-1$. We note that (\ref{masssin})
is consistent with the transformation behaviour of both
$\cFg{g}(z)$ and ${\cal M}(z)$ as holomorphic
sections. Indeed, it was shown in \cite{AGNT2} that the
leading holomorphic singularity of the ${\cal F}^{(g)}$ 
on the heterotic side is
given by
\beqa
{\cal F}^{(g>1)}(T)
=  -\frac{2}{\pi}  \left(\frac{a}{2}\right)^g 
\frac{B_{2g}}{2g(2g-2)}\frac{1}{[{\cal \mu}(T)]^{2(g-1)}} \;\;,
\label{sinfg}
\eeqa
where $\mu(T) = \frac{1}{\pi}(T-1)$.  The identification
of $\mu$ follows by comparing \cite{AGNT2}
the singularity of the holomorphic
prepotential ${\cal F}^{(0)} = -\frac{1}{2} ST^2 + \frac{1}{ \pi}
(T-1)^2 \log (T-1)$ with $2 Z_{c=1} =  \mu^2 \log \mu$.
The relative factor of $\frac{1}{\pi}\left(\frac{a}{2}\right)^g 
$ between (\ref{sinfg}) 
and (\ref{sinfgh})
reflects a different normalization convention adopted in this paper
from the one used in \cite{AGNT2}, whereas
the relative factor of $2$ 
reflects the fact that there are two additional vector 
multiplets becoming massless at $T=1$. Noting that $\frac{2}{T-1} 
\approx  \partial_T
\log (j(T) - j(1))$, where $j$  denotes
the modular invariant function $j=E_4^3/\eta^{24}$, 
it follows that
the leading singularity of
${\cal F}^{(g)}$ can be rewritten into the 
following holomorphic manifestly modular covariant
form with the correct modular weight 
\beqa
{\cal F}^{(g>1)}(T)
=  \frac{1}{\pi} \left(\frac{a}{2}\right)^g 
\frac{B_{2g}}{2g(2g-2)(2g-2)!}
{\hat D}_T^{2(g-1)} \log(j(T) - j(1)) \;\;,
\label{leadst}
\eeqa
where the holomorphic modular covariant derivative ${\hat D}_T$ of a 
weight-$\o$ modular form $P(T)$ is given by \cite{CFILQ}
\beqa
{\hat D}_T P(T)=\Big(\partial_T - \o \,G_2(T)\Big)P(T) \;\;,
\label{modcd} 
\eeqa
and where $G_2(T)$ is related to the Dedekind function $\eta(T)$ by
$G_2(T)= \pa_T \log \eta^2(T)$. For later use we define the
non-holomorphic modular function $\hat G_2(T)$ of weight 2,
\beq
\hat G_2(T,\bar T)= {1\over T+\bar T} +  G_2(T)  \,. \label{hatG2}
\eeq
For comparison, we also give the K\"ahler covariant derivative 
for a section $P(T)$ that transforms under
projective transformations with weight $w$. It reads
\beq
D_TP(T)\equiv \Big(\pa_T + w\,\pa_T \hat K\Big) P(T) = \Big(\pa_T -
{w\over T+\bar T}\Big)P(T)\,.
\eeq
As the projective and the modular weights are opposite, $w=-\o$
so that the K\"ahler covariant derivative and the holomorphic
covariant derivatives differ by a covariant term equal to
$\o\,\hat G_2(T,\bar T)\,P(T)$. 

Next, let us investigate the leading  holomorphic singularities for
the heterotic model with three vector fields $S$, $T$ and
$U$. In the same way as above, we find the perturbative expression
for the holomorphic mass for elementary string states 
${\cal M}$ for this $S$-$T$-$U$ model,
\beqa
{\cal M}(T,U) = M_0 + M_1 \,T U  + i M_2 \,T + i M_3 \,U \;\;.
\label{hmass}
\eeqa
The holomorphic mass (\ref{hmass}) can vanish at certain lines/points
in the perturbative moduli space, namely at $T=U \neq 1,
e^{\frac{i \pi}{6}}$, 
at $T=U=1$ and at $T=U= e^{\frac{i \pi}{6}}$.  At these lines/points, the 
$U(1)^2$ associated with the $T$ and the $U$ moduli get enhanced
to $U(1) \times SU(2)$,  $SU(2)^2$ and $SU(3)$, respectively.
Hence, the number of additional
vector multiplets at these lines/points is two, four and six, respectively.
As before, one expects the leading holomorphic singularities to be
given by (\ref{masssin}). More precisely, 
in the chamber $T > U$, one expects that, as $T \rightarrow U$,
\beqa
{\cal F}^{(g)}(T,U)
\propto  - 2 \left(\frac{1}{T-U}\right)^{2(g-1)}\;\;,  
\label{sin1stu}
\eeqa
whereas as $T \rightarrow U=1$
\beqa
{\cal F}^{(g)}(T,U)
\propto  - 4 \left(\frac{1}{T-1}\right)^{2(g-1)} \;\;,
\label{sin2stu}
\eeqa
and finally, as $T \rightarrow U=\rho=e^{\frac{i \pi}{6}}$,
\beqa
{\cal F}^{(g)}(T,U)
\propto  - 6 \left(\frac{1}{T-\rho}\right)^{2(g-1)}\;\;.
\label{sin3stu}
\eeqa
The unique holomorphic modular covariant generalisation
of (\ref{sin1stu}) transforming 
as in (\ref{mwstu}) under modular transformations, is as follows
(in the chamber $T > U$) 
\beqa
{\cal F}^{(g>1)}(T,U)
=  \beta_g\, 
{\hat D}_T^{g-1} {\hat D}_U^{g-1}\log(j(T) - j(U)) \;\;,
\label{leadstu}
\eeqa
where 
\beq
\beta_g= -  \frac{1}{\pi} \left(\frac{a}{2}\right)^g 
\frac{B_{2g}}{g(2g-2)(2g-2)!}(-)^{g} \;.\quad (g>1) 
\eeq
The relative factor of $2$ between (\ref{leadstu}) and (\ref{leadst})
is a reflection of the fact that in the $S$-$T$-$U$ model twice as 
many states as in the $S$-$T$ model become massless at $T=1$ .

Finally, it should be pointed out that there can also be
subleading holomorphic singularities in the Wilsonian couplings
${\cal F}^{(g)}$ as well as subleading 
non-holomorphic singularities in the full non-holomorphic
couplings $\cFgc{g}$.  Clearly, the latter ones are uniquely
determined by the anomaly equations, as we will show in the 
remaining subsections for the case of $\cFgc{2}$ and 
$\cFgc{3}$ in the $S$-$T$-$U$ model.


\subsection{The modular covariant section $\cFgc{2}$ 
in the $S$-$T$-$U$ model}

In this section, we will solve the holomorphic anomaly equations 
(\ref{f2het}) for  $\cFgc{2}$.
First consider the non--holomorphic $\cFgc{1}$
in the weak-coupling limit $\modS \rightarrow \infty$.
In the chamber
$T>U$ \cite{CLM,Curio,CLM2},
\beqa
\cFgc{1} = a S_{\rm inv} + h_{\rm inv}^{(1){\rm cov}} \;, 
\label{f1cov}
\eeqa
where $h^{(1){\rm cov}}_{\rm inv}$ decomposes into two terms,
\beqa
h_{\rm inv}^{(1){\rm cov}} = 
\frac{b_{\rm grav}}{2 \pi}
\left(\hat{K}(T,{\bar T},U,{\bar U}) + \log
  \eta^{-2}(T)\,\eta^{-2}(U) \right) + 
\b_1\, \log(j(T) - j(U))\,.
\label{hinv}
\eeqa
Note that $\cFgc{1}$ solves the heterotic version of the
anomaly equation (\ref{nonholoF1}). The second term is modular invariant, 
as is the real part of the first term. However, holomorphic 
derivatives of $h^{(1)\,\rm cov}_{\rm inv}$ constitute modular forms, and
they are the quantities that will play a role below. Furthermore, 
$b_{\rm grav}$
denotes the gravitational beta function.
In the standard normalization \cite{AL}, $a=24$ and 
$b_{\rm grav}=528$. The last term proportional to $\b_1$
represents the holomorphic singularity of $\cFgc{1}$ 
\cite{CLM,Curio}, which is not covered by the arguments of the
previous section. The coefficient $\b_1$ equals
\beq
\b_1 = \frac{2}{ \pi}\,.
\eeq

Since the anomaly equation
(\ref{f2het}) and the one that follows from it by interchanging
$T$ and $U$ are linear, we can solve them term by term. We  distinguish 
three different terms. The first one requires to solve the equation 
\beq 
[{\partial}_{\bar T} \cFgc{2}]_1  =  
\frac{a^2}{(T + \bar T)^2}\,\partial_U   V_{\rm inv} 
\;, \qquad 
[{\partial}_{\bar U} \cFgc{2}]_1  = 
 \frac{a^2}{(U + \bar U)^2} \,\partial_T V_{\rm inv} \,,
\label{Avin}
\eeq
where $V_{\rm inv} =   V - \sigma - \bar{\sigma}$, as before,
with \cite{DWKLL}
\beqa
\sigma(T,U) =  - {\textstyle\frac{1}{2}} \partial_T \partial_U h
+ {\textstyle {1\over 8}} L(T,U) \;, \quad {\rm with} \quad
L(T,U)=- \frac{4}{\pi} \log (j(T) - j(U)) \;. 
\label{svinv}
\eeqa
We recall that $V_{\rm inv}$
is not only invariant under target--space duality transformations, but
also finite everywhere inside the perturbative moduli space.

The integrability of (\ref{Avin}) is easily shown by using the following
identities, which are special cases of (\ref{GSidentity}), 
\beqa
\partial_{\bar T} \partial_{T} { V} = \frac{2V
  +\pa_T\pa_Uh+\pa_{\bar T}\pa_{\bar U} \bar h}{(T + {\bar T})^2}
\,, \;\;\;\;\;\;
\partial_{\bar U} \partial_{U} V = \frac{ 2V
  +\pa_T\pa_Uh+\pa_{\bar T}\pa_{\bar U} \bar h }{(U + {\bar
    U})^2}   \,.
\label{integr}
\eeqa
Here we have used the explicit form of $V$ given in (\ref{GSfunction}).
Note that, although 
 the numerator is modular invariant, it is singular at
the lines/points of semi-classical gauge--symmetry enhancement. The
above integrability relation is to be expected, as we know
that the full anomaly equations are integrable from the very
beginning, as we discussed in section~3.

By making use of (\ref{integr}), it follows that the following ansatz for 
$\cFgc{2}$, 
\beqa
[\cFgc{2}]_1 = {\textstyle\frac{1}{2}}a^2\, \partial_T \partial_U
V_{\rm inv} 
+ {\textstyle{1\over 8}} a^2 \left( {\hat G}_2(T,{\bar T})\,
  \partial_U L +{\hat G}_2(U,{\bar U}) \,\partial_T L \right)\,,
\label{f2acov}
\eeqa
solves (\ref{Avin}). The modular form $\hat G_2$ was defined in
(\ref{hatG2}). Note that $\cFgc{2}$, given in (\ref{f2acov}),
has the appropriate modular weight  and that it also does not
contribute to the leading
holomorphic singularity given in (\ref{leadstu}).   Here,
we have made use of the freedom of adding holomorphic terms to $\cFgc{2}$
in order to arrive at (\ref{f2acov}).

Next, consider solving the second and the third term of the anomaly equation,
which correspond to solving
\beqa 
[{\partial}_{\bar T} \cFgc{2}]_{2+3}  =
-\frac{a}{(T + \bar T)^2} \,
\partial_U  h_{\rm inv}^{(1){\rm cov}}
\;,\quad 
[{\partial}_{\bar U} \cFgc{2}]_{2+3}  = 
- \frac{a}{(U + \bar U)^2}\,
\partial_T h_{\rm inv}^{(1){\rm cov}}\,,
\label{B}
\eeqa
where $h_{\rm inv}^{(1){\rm cov}}$ is given in (\ref{hinv}).
Consider the first term in $h_{\rm inv}^{(1){\rm cov}}$, 
\beqa 
[{\partial}_{\bar T} \cFgc{2}]_2  &=&  
- \frac{a\, b_{\rm grav}}{2 \pi}\frac{1}{(T + \bar T)^2} 
\partial_U  
\Big(\hat{K}(T,{\bar T},U,{\bar U}) + \log \eta^{-2}(T)\,\eta^{-2}(U) 
\Big) 
\,,\nonumber\\
{}[{\partial}_{\bar U} \cFgc{2}]_2  
&=& -  \frac{a\,b_{\rm grav}}{2 \pi}
\frac{1}{(U + \bar U)^2}\partial_T \Big(\hat{K}(T,{\bar T},U,{\bar U}) 
+ \log \eta^{-2}(T)\,\eta^{-2}(U) \Big)\,,
\label{kahler}
\eeqa
which is solved by the modular covariant expression
\beqa
[\cFgc{2}]_2 = -  \frac{a\, b_{\rm grav}}{2 \pi}
 \hat{G}_2 (T,{\bar T})\, \hat{G}_2 (U,{\bar U}) \,.
\label{f2b}
\eeqa
The above expression (\ref{f2b}) was also recently derived in
\cite{AGNT3} in the context of heterotic $N=1$ string vacua. 

Then consider the second term of $h_{\rm inv}^{(1){\rm cov}}$,  
\beq
[{\partial}_{\bar T} \cFgc{2}]_3  = 
- \frac{a}{(T + \bar T)^2} \partial_U  \cFg{1} \;, \qquad 
[{\partial}_{\bar U} \cFgc{2}]_3  = - \frac{a}{(U + \bar
  U)^2}\partial_T \cFg{1},
\label{diffj}
\eeq
where 
\beqa
\cFg{1} = \b_1\, \log(j(T) - j(U)).
\label{logj}
\eeqa

One can proceed in two ways.  One can either solve
(\ref{diffj}) directly, or one can use the symplectic formalism developed
in section \ref{shanom} 
in order to construct a modular covariant non-holomorphic
solution $\cFgc{2}$ to the anomaly equations (\ref{diffj}).  
This is so, because the construction of 
symplectic functions  (\ref{Fg}) is based on the existence of a 
holomorphic section $\cFg{1}$, and the $\cFg{1}$ given in (\ref{logj})
is precisely such an object.  
Thus, we will use the latter strategy in the following.

The explicit expression for $\cFgc{2}$,
which one obtains from (\ref{Fg}) for the
$S$-$T$-$U$ model, 
is derived in appendix \ref{fgstu}.  
In the limit $\modS\to\infty$, it is given by
\beqa
\cFgc{2}(T,U,{\bar T},{\bar U})
 = \cFg{2} (T,U) + \frac{a}{T + \bar{T}} \partial_U {\cal
   F}^{(1)} +  \frac{a}{U + \bar{U}} 
\partial_T {\cal F}^{(1)}  \;\;,
\label{f2nh}
\eeqa
where the holomorphic $\cFg{1}$ is given by (\ref{logj}).
Note that the non-holomorphic part of (\ref{f2nh}) is not
modular covariant.  Just
as before, in order to obtain
a target--space duality covariant expression for $\cFgc{2}$,
one has to include an appropriate {\it holomorphic} $\cFg{2} (T,U)$ in
(\ref{f2nh}). 
One such appropriate $\cFg{2} (T,U)$ is given by
\beqa
{\cal F}^{(2)} (T,U) = a \,G_2(T)\, \partial_U {\cal F}^{(1)}
+ a \, G_2(U) \,\partial_T {\cal F}^{(1)} + \beta_2\,
\partial_T \partial_U\log(j(T) - j(U)) \;\;,
\label{holf2}
\eeqa
where $G_2$ was introduced in (\ref{modcd}).
The last term in (\ref{holf2}) is the leading holomorphic singularity
(\ref{leadstu}).
Combining (\ref{holf2}) with (\ref{f2nh}) yields
\beqa
[\cFgc{2}]_3 = a \left(\hat{G}_2(T,{\bar T}) \,\partial_U \cFg{1} + 
\hat{G}_2(U,{\bar U})\, \partial_T\cFg{1} \right) + \beta_2\,
\partial_T \partial_U\log(j(T) - j(U))\,.
\label{f2tu}
\eeqa
Expression (\ref{f2tu}) is manifestly modular covariant, and
it can be checked in a straightforward way that it solves 
the anomaly equation (\ref{diffj}).

Thus, the solution $\cFgc{2}$ to the anomaly equations (\ref{f2het})
is given by the sum of (\ref{f2acov}), (\ref{f2b}) and (\ref{f2tu}), 
that is by
\beqa
\cFgc{2} &=& {\textstyle\frac{1}{2}}a^2\, \partial_T \partial_U
V_{\rm inv} 
+ {\textstyle{1\over8}} a^2 \left( {\hat G}_2(T,{\bar T})\, \partial_U L + 
{\hat G}_2(U,{\bar U})\, \partial_T L \right) \nonumber\\
&&- \frac{a\,b_{\rm grav}}{2 \pi}
 \hat{G}_2 (T,{\bar T})\, \hat{G}_2 (U,{\bar U})  \label{fullf2cov}\\
&&+ a \left(\hat{G}_2(T,{\bar T})\, \partial_U \cFg{1} + 
\hat{G}_2(U,{\bar U})\, \partial_T\cFg{1} \right) + \beta_2\,
\partial_T \partial_U\log(j(T) - j(U)). \nonumber
\eeqa
Note that there is no freedom left in adding further holomorphic 
modular forms to (\ref{fullf2cov}). The reason is that 
target-space duality and the 
knowledge of the leading holomorphic 
singularities that are associated with known 
gauge--symmetry enhancement points/lines,  fixes the structure of $\cFgc{2}$ 
completely.  Thus, the $\cFgc{2}$ given in (\ref{fullf2cov}) is the 
full solution to (\ref{f2het}).  
In appendix \ref{hyperelliptic}, we will give the power-series 
expansion of $\cFgc{2}(T,{\bar T},U,{\bar U})$ 
in the limit ${\bar T}, {\bar U} \rightarrow
\infty$.


\subsection{The modular covariant section $\cFgc{3}$
  in the $S$-$T$-$U$ model}

In this section, we will solve the holomorphic anomaly equations 
(\ref{fghetweak})
for  $\cFgc{3}$.  In order to solve
them, we will need to evaluate $D_U \partial_T \partial_U V_{\rm inv}$.
We find that
\beqa
D_U \partial_T \partial_U V_{\rm inv}
= {\textstyle\frac{1}{2}} D_T^2 \partial^3_U h -{\textstyle
\frac{1}{8}}  D_U \partial_T \partial_U L
\;.
\label{dddvinv}
\eeqa
We recall that the term $\partial_U^3 h$ appearing on the right hand side
of (\ref{dddvinv}) transforms covariantly under target--space 
duality transformations, and that it has modular weights $-2$ and $4$
under $SL(2,{\bf Z})_T$ and 
$SL(2,{\bf Z})_U$ transformations, respectively \cite{DWKLL}.
Thus, both terms appearing on the 
right hand side of (\ref{dddvinv}) have modular weights $2$ and $4$
under $SL(2,{\bf Z})_T$ and $SL(2,{\bf Z})_U$ transformations.

We will now solve the anomaly equations (\ref{fghetweak}) for $\cFgc{3}$
using (\ref{fullf2cov}) as the input on the right hand side of 
(\ref{fghetweak}).
Since these anomaly equations are
linear, we will solve them separately
for each of the three lines of (\ref{fullf2cov}).
First consider the anomaly equations based on the first line of 
(\ref{fullf2cov}).  Using (\ref{dddvinv}), it follows that these 
differential equations can be written out as follows
\beqa
[\partial_{\bar T} \cFgc{3}]_1 &=& - \frac{a^3}{(T + {\bar T})^2} \Big[
{\textstyle\frac{1}{4}} D_T^2 \partial_U^3 h (T,U) -
{\textstyle\frac{1}{16}} D_U \partial_T \partial_U L(T,U)  \label{f3v} \\
&&\hspace{20mm} +{\textstyle \frac{1}{8}} D_U\Big({\hat G}_2 (U , {\bar U})\,
\partial_T L(T,U)  +
{\hat G}_2 (T , {\bar T}) \, \partial_U L(T,U)\Big)\Big] \;,
\nonumber
\eeqa
and likewise with $T$ and $U$ interchanged.  Note that 
(\ref{f3v}) is modular covariant.
It can be solved
in a rather straightforward way, and 
the solution to (\ref{f3v}) reads
\beqa
[\cFgc{3}]_1 &=& {\textstyle\frac{1}{4}}a^3 \Big[
 {\hat G}_2(T,{\bar T})\, D_T^2  \partial_U^3 h + 
{\hat G}^2_2(T,{\bar T}) \,D_T
\partial_U^3 h + {\textstyle\frac{2}{3}} {\hat G}^3_2(T,{\bar T})
\, \partial_U^3 h   \nonumber\\
 &&\hspace{10mm} +  {\hat G}_2(U,{\bar U})\, D_U^2  \partial_T^3 h + 
{\hat G}^2_2(U,{\bar U}) \, D_U
\partial_T^3 h + {\textstyle\frac{2}{3}} {\hat G}^3_2(U,{\bar U})
\,\partial_T^3 h \Big]  \nonumber\\
&&-{\textstyle\frac{1}{16}}a^3 \Big[ {\hat G}_2(T,{\bar T})\,
  D_U \partial_T \partial_U L 
+ {\hat G}_2(U,{\bar U}) \,D_T \partial_T \partial_U L \nonumber\\ 
&&\hspace{10mm} - 2 {\hat G}_2(T,{\bar T})\,{\hat G}_2(U,{\bar U})\, 
\partial_T \partial_U L  \Big] \nonumber\\
&&+ {\textstyle\frac{1}{8}} a^3
\Big[{\hat G}_2(T,{\bar T}) \,D_U{\hat G}_2(U,{\bar U})\,
  \partial_T L +
{\hat G}_2(U,{\bar U}) \,D_T {\hat G}_2(T,{\bar T})\, \partial_U L 
\nonumber\\
&& \hspace{10mm} + {\textstyle\frac{1}{2}} {\hat G}^2_2 (T,{\bar
  T}) \, D_U \partial_U L +
{\textstyle\frac{1}{2}} {\hat G}^2_2 (U,{\bar U}) \,D_T
\partial_T L \nonumber\\ 
&&\hspace{10mm}-   {\hat G}^2_2 (U,{\bar U})\, {\hat G}_2
  (T,{\bar T}) \,\partial_T L 
- {\hat G}^2_2 (T,{\bar T})\, {\hat G}_2 (U,{\bar U})\, \partial_U L
\Big] \;.
\label{f3a}
\eeqa
Next, consider solving (\ref{fghetweak}) based on the second line
of (\ref{fullf2cov}).  The solution reads
\beqa
[\cFgc{3}]_2 &=& - \frac{a^2\, b_{\rm grav}}{4 \pi} \left[ {\hat
    G}_2^2(T,{\bar T})\, D_U {\hat G}_2(U,{\bar U}) + 
{\hat G}_2^2(U,{\bar U})\,
D_T {\hat G}_2(T,{\bar T}) \right. \nonumber\\
&&\hspace{15mm}-  \left. {\hat G}_2^2 (T,{\bar T})\,
{\hat G}_2^2 (U,{\bar U}) \right]\, . 
\label{f3b}
\eeqa
Note that one can in principle
add further modular covariant holomorphic terms to either (\ref{f3a}) or
(\ref{f3b}), such as $[G_2^2(T) - \partial_T G_2(T)]
[G_2^2(U) - \partial_U G_2(U)]$, for example \cite{BCOV}.

Finally, consider solving (\ref{fghetweak}) 
based on the third line of (\ref{fullf2cov}). One can again use
the symplectic formalism developed in section \ref{shanom} 
in order to construct a modular covariant non-holomorphic
solution $\cFgc{3}$.  
The explicit expressions for  $\cFgc{3}$,
which one obtains from (\ref{Fg}) for the $S$-$T$-$U$ model, 
are presented in appendix \ref{fgstu}.  
They are given by 
\beqa
\cFgc{3} (T,U,{\bar T},{\bar U})
&=& {\cal F}^{(3)} (T,U) + a \bigg[
\frac{\partial_T {\cal F}^{(2)} }{U + \bar{U}} 
+ \frac{\partial_U  {\cal F}^{(2)}}{T+\bar{T}} 
+ \frac{2{\cal F}^{(2)}}{(T + \bar{T})(U+\bar{U})}  \bigg] \nonumber\\
&&+ {\textstyle{1\over2}} a^2 \bigg[
 \frac{1}{(T + \bar{T})^2} \partial_U^2 {\cal F}^{(1)}
+ 
 \frac{1}{(U + \bar{U})^2} \partial_T^2 {\cal F}^{(1)} \nonumber\\
&& \hspace{12mm} +
\frac{2}{(T+\bar{T})(U + \bar{U})} \partial_T \partial_U 
{\cal F}^{(1 )} \bigg]  \label{nholf3} \\
&&+  \frac{a^2}{(T+\bar{T})(U + \bar{U})}\bigg[\frac{\partial_U 
{\cal F}^{(1)} }{T+\bar{T}} 
+
\frac{\partial_T {\cal F}^{(1)}}{U+\bar{U}}  \bigg] \,,
\nonumber 
\eeqa
where the holomorphic ${\cal F}^{(1)}$ and ${\cal F}^{(2)}$ 
are given in equations
(\ref{logj}) and (\ref{holf2}), respectively.
Again, note that the non-holomorphic part 
of $\cFgc{3}$ is not modular covariant.  As before, in order to obtain
a target--space duality covariant expression for $\cFgc{3}$,
one has to include an appropriate holomorphic $\cFg{3}(T,U)$ in (\ref{nholf3}).
One choice is as follows 
\beqa
{\cal F}^{(3)} (T,U)\! &=\!& a^2 \Big[
{\textstyle\frac{1}{2}} G_2^2(T) \,\partial_U^2 {\cal F}^{(1)} + 
{\textstyle\frac{1}{2}} G_2^2(U) \partial_T^2 {\cal F}^{(1)}
  \nonumber\\
&&\hspace{10mm}
- \Big(G_2(U)\, G_2^2(T) - \partial_T G_2(T) \,G_2(U)\Big)
\partial_U {\cal F}^{(1)}
 \nonumber\\
&&\hspace{10mm}-
\Big( G_2(T) \,G_2^2(U) - \partial_U G_2(U)\, G_2(T)\Big)
\partial_T {\cal F}^{(1)}
+ G_2(T)\, G_2(U)\, \partial_T \partial_U {\cal F}^{(1)} \Big] \nonumber\\
&&- a \beta_2 \Big[
G_2(T) \,G_2(U)\, \partial_T \partial_U \log(j(T)-j(U))  \nonumber\\
&&\hspace{10mm}-G_2(T) \,\partial_T \partial_U^2 \log(j(T)-j(U))
-  G_2(U)\, \partial_U \partial_T^2 \log(j(T)-j(U))\Big] \nonumber\\
&&+ \beta_3
{\hat D}_T^{2} {\hat D}_U^{2}\log(j(T) - j(U)) \;.
\label{hf3}
\eeqa
The last term in (\ref{hf3}) is the leading holomorphic singularity
(\ref{leadstu}).  Inserting (\ref{hf3}) into (\ref{nholf3})
yields
\beqa
[\cFgc{3}]_3 \!&=\!&  a^2 \bigg[{\textstyle \frac{1}{2} }
{\hat G}^2_2(T) \,D^2_U {\cal F}^{(1)} + 
D_U \left({\hat G}_2(U)\, {\hat G}_2(T)\, D_T {\cal F}^{(1)} \right)
 \nonumber\\
&&\hspace{10mm}+{\textstyle \frac{1}{2} }
{\hat G}^2_2(U) \,D^2_T {\cal F}^{(1)} + 
D_T\left( {\hat G}_2(T) \,{\hat G}_2(U)\, D_U {\cal F}^{(1)} \right) \bigg]
\left. \right. \nonumber\\
&&+ a \beta_2 \bigg[
{\hat G}_2 (T)\, D_T D^2_U \log(j(T) - j(U)) + {\hat G}_2 (U)\, 
D_U D^2_T \log(j(T) - j(U))
\nonumber\\
&&\hspace{10mm}- 2  {\hat G}_2(T)\, 
{\hat G}_2(U)\, D_U D_T \log(j(T) - j(U)) \bigg] \nonumber\\
&&- a^2 \bigg[ {\hat G}_2(T)\, {\hat G}_2(U) \,D_U D_T {\cal F}^{(1)} 
+ {\hat G}_2(T)\, {\hat G}^2_2(U) \,D_T {\cal F}^{(1)} \nonumber\\
&&\hspace{10mm}+ {\hat G}_2(U) \,{\hat G}^2_2(T) \,D_U {\cal F}^{(1)} \bigg]
+  \beta_3
{\hat D}_T^{2} {\hat D}_U^{2}\log(j(T) - j(U)) \;.
\label{f3final}
\eeqa
The expression (\ref{f3final}) is modular covariant, and again
it can be checked that it solves the holomorphic
anomaly equation (\ref{fghetweak}) based on (\ref{f2tu}).
In principle, one could again add additional holomorphic
modular covariant terms of the correct modular weight
to (\ref{f3final}).  Such additional terms could, for instance, describe
holomorphic subleading singularities.  An example of such a 
singular subleading term is $[G_2^2(T) - \partial_T G_2(T)]
[G_2^2(U) - \partial_T G_2(U)] \log(j(T) - j(U))$.  

Thus, the full solution $\cFgc{3}$ (up to possible additional
holomorphic modular forms of weight 4) to the anomaly equation
(\ref{fghetweak}) is given by the sum of (\ref{f3a}), (\ref{f3b}) and
(\ref{f3final}).


\section{Conclusions}
In this paper we have discussed the computation of the 
moduli-dependent, higher-order gravitational couplings 
in the perturbative limit
of four-dimensional heterotic string compactifications with $N=2$
space--time supersymmetry. Our method of calculating the couplings
$\cF^{(g)}$ consisted in solving the anomaly equation 
(\ref{anomsbig}), where we took as an input the known one-loop
expressions for the heterotic prepotential $\cF^{(0)}$ and
the gravitational couplings $\cF^{(1)}$, as well as the
leading holomorphic singularity structure of the higher $\cF^{(g)}$.
Subsequently we imposed target--space duality covariance. 
For the $S$-$T$-$U$ model the result can thus be explicitly written in 
terms of modular forms. The anomaly equation
can be derived, either from string--string duality by taking 
the type-II anomaly 
equation (\ref{anomalyeq}) in the limit of a large K\"ahler-class 
modulus, so as to make contact with the perturbative heterotic side,
or by exploiting arguments based on symplectic covariance. The
results of these two approaches coincide and lead to the 
anomaly equation (\ref{anomsbig}).

Let us briefly recall the relevant steps in the derivation of 
 $\cFgc{2}$, given in (\ref{fullf2cov}), and of $\cFgc{3}$, given
as the sum of eqs. (\ref{f3a}), (\ref{f3b}) and (\ref{f3final}),
for the $S$-$T$-$U$ model.
As already emphasized, the one-loop 
$\cF^{(1)}$ is both holomorphic and duality invariant
near the region $T=U$ in the moduli space.
We can therefore use the symplectic formalism to compute the corresponding
non-holomophic part
of $\cF^{(2)}$. Adding an appropriate holomorphic function yields a
covariant expression for $\cF^{(2)}$. Second, $\cF^{(1)}$ contains
further terms which are non-singular in the limit $T\rightarrow U$,
namely a one-loop Green--Schwarz piece 
from the invariant dilaton and a term proportional
to $b_{\rm grav}$, related to the one-loop K\"ahler and $\sigma$-model
anomalies. In order to obtain the covariant $\cFgc{2}$ that belongs
to these non-singular terms, we explicitly solved the holomorphic
anomaly equation. For the  higher $\cF^{(g)}$, this procedure can be
continued. Those terms which are derived from the holomorphic and invariant
$\cF^{(1)}$ are constructed by using the covariant 
derivative (\ref{derivative}).
The remaining terms, which are obtained from the Green--Schwarz term
and from the term proportional to $b_{\rm grav}$ 
in $\cF^{(1)}$, follow from the holomorphic anomaly equation.

In summary, we have found a very transparent and systematic structure
in the process of solving the anomaly equation on the heterotic side
and in the
form of its solutions.  Then, in principle,
important information about the higher-genus instanton
numbers of the relevant dual Calabi--Yau three-fold can be obtained
through the second-quantized mirror map
\cite{FerHarStrVa}.

\vspace{1cm}

{\bf Acknowledgement}\\[3mm]
We would like to thank G. Curio and J. Louis for useful discussions.
S.--J.R. was supported in part by U.S.NSF-KOSEF Bilateral Grant, 
KOSEF Purpose-Oriented Research Grant 94-1400-04-01-3 and SRC-Program,
KRF Non-Directed Research Grant 81500-1341 and International Collaboration 
Grant, and Ministry of Education BSRI Grant 94-2418. The work of T. M. was
supported by DFG.

This work was carried out during the preparation
of the TMR program ERB-4061-PL-95-0789 in which B.~d.W. is associated
to Utrecht and D.~L. and T.~M. are associated to Berlin.

\appendix


\section{Some asymptotic results for $\modS\to\infty$} \label{sbig}
\setcounter{equation}{0}

We wish to consider certain expressions in the limit that
$\modS\to\infty$, based on the class of functions (\ref{cFnull}).
We expand the corresponding K\"ahler potential (\ref{Kpotential})
as
\beq
K(S,\bar S,T,\bar T) \approx -\log (S+\bar S)+ \hat K(T,\bar T) - 
{V(T,\bar T)\over S+\bar S}  + \cdots\,,
\eeq
where the ellipses denote terms of higher order in $(\modS)^{-1}$.
$\hat K$ and the Green-Schwarz term $V$ were already given in
(\ref{hatkahler}) and (\ref{defGS}). 

First consider the components of the metric in the large-$(\modS)$ limit,
\beqa
g_{A\bar B}&=& 
  \pmatrix{ {\strut\displaystyle 1\over \strut\displaystyle(S+\bar
   S)^2}-{\strut\displaystyle 2V\over\strut \displaystyle(\modS)^3} &
      {\strut\displaystyle V_{\bar b}\over\strut\displaystyle
        (\modS)^2}\cr 
\noalign{\vskip5mm} 
  {\textstyle\strut\displaystyle V_a\over\strut\displaystyle
    (\modS)^2}& \hat g_{a\bar b}- {\strut\displaystyle
    V^{a\bar b}\over\strut\displaystyle \modS}\cr} + \cdots \,, 
\nonumber \\[5mm] 
g^{A\bar B}&=& \pmatrix{(\modS)^2+ 2(\modS)V& - V^{\bar b}\cr
\noalign{\vskip5mm}
 - V^a &\hat g^{a\bar b}+ {\strut\displaystyle V^{a\bar
      b}\over\strut\displaystyle  \modS}\cr}+
\cdots\,. \label{metricsbig} 
\eeqa
Here $\hat g_{a\bar b}$ is the metric associated with the K\"ahler 
potential $\hat K(T,\bar T)$,
which is used to raise and lower indices in the above formula. The 
explicit expressions are shown in (\ref{hatkahler}). 
Sub- and superscripts $a, b,\ldots$
and $\bar a ,\bar b,\ldots$ attached to $V$ denote
differentiations of $V$ with respect to $T^a,T^b,\ldots$ and
$\bar T^a,\bar T^b,\ldots$ that are covariant under diffeomorphisms
in the K\"ahler space parametrized by these coordinates. Although
these covariantizations do not contribute to the expressions
above, they will contribute to some of the formulae below. 

Similarly we evaluate the following expressions for large
$\modS$, without making any assumptions on the large-$(\modS)$
behaviour of the functions $\cF_1$, $\cF_2$ and $\cF$ on which
the various derivatives act, 
\beqa
e^{2K}\, D^{\bar S}{\cal F}_1\,D^{\bar S}\cF_2 &\approx&
e^{2\hat K} (\modS)^2  \Big(\cD_S\cF_1 - {V^a\over (\modS)^2}\cD_a
  \cF_1\Big)\,\Big(\cD_S\cF_2 - {V^a\over (\modS)^2}\cD_a
  \cF_2\Big)\,, \nonumber \\ 
e^{2K} \,D^{\bar a}{\cal F}_1\,D^{\bar S}\cF_2 &\approx& e^{2\hat
  K} \left(V^{\bar 
    a}\cD_S\cF_1 - \cD^{\bar a} \cF_1\right) \, \Big(-\cD_S\cF_2
  + {V^a\over (\modS)^2}\cD_a \cF_2\Big)\,, \nonumber \\
e^{2K} \,D^{\bar a}{\cal F}_1\,D^{\bar b}\cF_2 &\approx& {
  e^{2\hat K}\over (\modS)^2} \left(V^{\bar
    a}\cD_S\cF_1 - \cD^{\bar a} \cF_1\right) \,\left(V^{\bar
    a}\cD_S\cF_2 - \cD^{\bar a} \cF_2\right) \,,\label{DFDF}
\eeqa
and
\beqa
e^{2K} \,D^{\bar S}\,D^{\bar S}\cF &\approx& e^{2\hat K} \Big[(\modS)^2
\cD_S^2  - 2V^a\,\cD_S\cD_a \nonumber \\
&&\hspace{8mm} +{V^aV^b\over (\modS)^2}  \cD_a\cD_b + 2(\modS)
\cD_S +{{\cal V}^{a}\over (\modS)^2} \cD_a \Big]\cF\,, \nonumber \\ 
e^{2K} \,D^{\bar a}D^{\bar S}\cF &\approx&  e^{2\hat K}\Big[ -V^{\bar
    a}\cD_S^2 +\cD^{\bar a}\cD_S  + {V^{\bar a}
    V^a\over (\modS)^2} \cD_S \cD_a \nonumber \\
&&\hspace{8mm}-{V^a\over
    (\modS)^2}\, \cD^{\bar a}\cD_a +{{\cal V}^{\bar a}\over
    (\modS)^2} \cD_S 
  - { V^{\bar a a}\over (\modS)^2} \cD_a \Big] \cF\,, \nonumber  \\
e^{2K} \,D^{\bar a}D^{\bar b}\cF &\approx& { e^{2\hat K}\over
  (\modS)^2} \Big[ V^{\bar
    a}V^{\bar b}\cD_S^2 -(V^{\bar a} \cD^{\bar b}+V^{\bar b}
  \cD^{\bar a})\cD_S  \nonumber \\
&&\hspace{20mm} + \cD^{\bar a} \cD^{\bar b}
  - {V^{\bar a \bar b} \over (\modS)^2} \cD_S\Big]\cF\,, \label{DDF}
\eeqa
where ${\cal V}^a$ and  ${\cal V}^{\bar a}$ are more complicated
objects quadratic in $V$ and its derivatives. 
The derivatives $\cD$ contain the Levi-Civita connection
associated with the K\"ahler metric $\hat g_{a\bar b}$ and a
K\"ahler connection proportional to $\pa_A K$.

Finally, consider the matrix $N_{IJ}$ and evaluate its leading
behaviour for large $\modS$. First we decompose 
\beq
N_{IJ}= (N_0)_{IJ} + (\D N)_{IJ}\,, 
\eeq
where the first term corresponds to the first term in
(\ref{cFnull}) and $\D N$ contains
the terms related to the function $h$. Its nonvanishing matrix
elements are
\beqa
(\D N)_{00}&=& - 2h +2 h_aT^a - h_{ab}T^aT^b + \{{\rm
  h.c.}\}\,, \nonumber \\ 
(\D N)_{0a}=(\D N)_{a0}&=& i h_a -i h_{ab}T^b + \{{\rm
  h.c.}\}\,, \nonumber \\ 
(\D N)_{ab}&=& h_{ab}+ \{{\rm h.c.}\}\,. \label{DeltaN}
\eeqa

Ignoring the contributions from the function $h(t)$, which we
will deal with later in perturbation theory, we first use
evaluate the matrix $n^{AB}$ introduced in section~3. It reads 
\beq
n^{AB} = \pmatrix{ 3(\modS) e^{\hat K} & -3 (T+\bar T)^b e^{\hat
    K}\cr 
\noalign{\vskip3mm}
 -3(T+\bar T)^a e^{\hat K} & {\strut\displaystyle -3 \over
   \strut\displaystyle \modS} \Big(\eta^{ab} -  (T+\bar
T)^a(T+\bar T)^b  e^{\hat K} \Big) \cr}\,.
\eeq
With the aid of this result and (\ref{inverseN}) we evaluate the
inverse of $(N_0)_{IJ}$, 
\beq
(N_0)^{IJ} = {e^{\hat K}\over \modS} \pmatrix{  2&i(S-\bar S) &
  i(T-\bar T)^b\cr
\noalign{\vskip3mm}
i(S-\bar S) & 2S\bar S & - S\, T^b -\bar S \,\bar T^b\cr
\noalign{\vskip3mm}
  i(T-\bar T)^a & - S\, T^a -\bar S \,\bar T^a & -{1\over 2} \eta^{ab}
  \,e^{-\hat K} +T^a \bar T^b + \bar T^a  T^b  \cr} \;\;. \label{inverseN0}
\eeq
Using (\ref{DeltaN}) and (\ref{inverseN0}) we can easily determine $N^{IJ}$ 
for large $\modS$.


\section{A class of non-holomorphic corrections to $\cFgc{2}$ and 
$\cFgc{3}$ in the $S$-$T$-$U$ model \label{fgstu}}
\setcounter{equation}{0}

In (\ref{thirdfc}) we gave the expressions for $\cFgc{2}$ and
$\cFgc{3}$ based on a general cubic function ${\cal F}^{(0)}$, in
terms of the holomorphic Wilson coefficient functions $\cFg{1}$,
$\cFg{2}$ and $\cFg{3}$. Here
we specialize these expressions to the case of the $S$-$T$-$U$
model. These sections are solutions of the truncated anomaly
equation (\ref{anomalyeq3}). At the end we exhibit their
behaviour for large $\modS$. 
We remind the reader that, according to (\ref{Fg}), $\cFgc{1}$
remains holomorphic.

Let us first give the K\"ahler potential for this case,
\beq
K(S,\bar S,T,\bar T,U,\bar U) = -\log (S+\bar S) -\log (T+\bar T)
-\log (U+\bar U) \,.
\eeq
Furthermore we list the explicit expressions for the inverse of the 
matrix $N_{IJ}$ defined in (\ref{defN}),
\beq
N^{IJ} =e^K 
\pmatrix{ 2&i (S-{ \bar{S}})&i (T-{ \bar{T}})&i(U-{ \bar{U}} )\cr 
\noalign{\medskip}
i(S-{\bar{S}})&2\,S{\bar{S}}& -ST-{ \bar{S}}\,{ \bar{T}}&
-SU-{\bar{S}}\,{ \bar{U}}\cr
\noalign{\medskip}
i (T-{\bar{T}} )&-ST-{\bar{S}}\,{ \bar{T}}&2\,T{ \bar{T}}&
-TU-{ \bar{T}}\,{ \bar{U}}\cr
\noalign{\medskip}i (U-{\bar{U}})&-SU-{\bar{S}}\,{\bar{U}}&
-TU-{\bar{T}}\,{\bar{U}}&2\,U{\bar{U}}\cr} \,,
\eeq
and the matrix $\wh{n}^{AB}$ defined in the text below (\ref{thirdfc}), 
\beq
\wh{n}^{AB} = - 6 \, e^K \; \pmatrix{
0 & (S+\bar{S}) (T + \bar{T}) & (S+\bar{S}) (U + \bar{U}) \cr
\noalign{\medskip}
(S+\bar{S}) (T + \bar{T}) & 0 & (T+\bar{T}) (U + \bar{U}) \cr
\noalign{\medskip}
(S+\bar{S}) (U + \bar{U}) &(T+\bar{T}) (U + \bar{U}) & 0 \cr}\,.
\eeq

With these definitions it is somewhat tedious but straightforward 
to calculate the expressions for $\cFgc{g}$ for this model, 
by using (\ref{thirdfc}). We remind the read that $\cFgc{1}$ remains 
equal to the holomorphic function $\cFg{1}$. 
The expression for $\cFgc{2}$ takes the form 
\beqa
{\cal F}^{(2)\,\ms{cov}} = {\cal F}^{(2)} +
\frac{1}{U+\bar{U}}\, {\cal F}^{(1)}_S\,{\cal F}^{(1)}_T\,
+\frac{1}{T+\bar{T}}\, {\cal F}^{(1)}_S\,{\cal F}^{(1)}_U\,
+\frac{1}{S + \bar{S}}\, {\cal F}^{(1)}_T\,
{\cal F}^{(1)}_U\,,
\eeqa
where the subscripts $S$, $T$ and $U$ denote ordinary partial 
derivatives with respect to these coordinates.
For large $\modS$ this reduces to (\ref{f2nh})
provided one makes use of (\ref{assum}).

The expression for $\cFgc{3}$ is much longer and reads 
\begin{eqnarray}
\cFgc{3}&\!=\!&  \cFg{3} + \frac{1}{U+\ov{U}} \, 
\left[ {\cal F}^{(2)}_S {\cal F}^{(1)}_T  + {\cal F}^{(2)}_T
  {\cal F}^{(1)}_S \right] \nonumber \\
&&
+ \frac{1}{T+\ov{T}} \, 
\left[ {\cal F}^{(2)}_U {\cal F}^{(1)}_S 
+ {\cal F}^{(2)}_S {\cal F}^{(1)}_U \right]
+\frac{1}{S+\bar{S}} \, 
\left[ {\cal F}^{(2)}_T {\cal F}^{(1)}_U 
+ {\cal F}^{(2)}_U {\cal F}^{(1)}_T \right]
\nonumber \\
 & & 
+\frac{2{\cal F}^{(2)}}{(S+\bar{S})(T+\bar{T})(U+\bar{U})}  
\left[ (S+\bar{S}) {\cal F}^{(1)}_S + (T+\bar{T}) {\cal F}^{(1)}_T 
+ (U+\bar{U}) {\cal F}^{(1)}_U \right]
\nonumber \\
&&
+ \frac{{\cal F}^{(1)}_{SS}} {2(T+\bar{T})^2 (U+\bar{U})^2} 
\left[ (T+\bar{T}) \, {\cal F}^{(1)}_T + (U+\bar{U}) {\cal
    F}^{(1)}_U \right]^2 
\nonumber \\
 & & 
+\frac{{\cal F}^{(1)}_{TT} }{2(U+\bar{U})^2 (S+\bar{S})^2} 
\left[ (U+\bar{U}) \, {\cal F}^{(1)}_U + (S+\bar{S}) {\cal
    F}^{(1)}_S \right]^2 
\nonumber \\
 & & 
+ \frac{{\cal F}^{(1)}_{UU} }{2(S+\bar{S})^2 (T+\bar{T})^2}
\left[ (S+\bar{S}) \, {\cal F}^{(1)}_S + (T+\bar{T}) {\cal
    F}^{(1)}_T \right]^2 
\nonumber \\
 & & 
+  \frac{{\cal F}^{(1)}_{ST}}{(S+\bar{S}) (T+\bar{T}) (U+\bar{U})^2 } 
\left[ (U+\bar{U})^2 ({\cal F}^{(1)}_U)^2 
+(U+\bar{U})(T+\bar{T}) \,{\cal F}^{(1)}_U  {\cal F}^{(1)}_T \right.
\nonumber \\
 & & 
\hspace{4cm}\left.
+(S+\bar{S})(U+\bar{U})\, {\cal F}^{(1)}_S  {\cal F}^{(1)}_U
+(T+\bar{T})(S+\bar{S})\, {\cal F}^{(1)}_T  {\cal F}^{(1)}_S
\right]
\nonumber \\
 & & 
+\frac{{\cal F}^{(1)}_{TU}}{(T+\bar{T}) (U+\bar{U}) (S+\bar{S})^2 }
\left[ (S+\bar{S})^2 ({\cal F}^{(1)}_S)^2 
+(U+\bar{U})(T+\bar{T})\, {\cal F}^{(1)}_U  {\cal F}^{(1)}_T \right.
\nonumber \\
 & & 
\hspace{4cm}\left.
+(S+\bar{S})(U+\bar{U})\, {\cal F}^{(1)}_S  {\cal F}^{(1)}_U
+(T+\bar{T})(S+\bar{S})\, {\cal F}^{(1)}_T  {\cal F}^{(1)}_S
\right]
\nonumber \\
 & & 
+ \frac{{\cal F}^{(1)}_{SU}}{(U+\bar{U}) (S+\bar{S}) (T+\bar{T})^2 } 
\left[ (T+\bar{T})^2 ({\cal F}^{(1)}_T)^2 
+(U+\bar{U})(T+\bar{T})\, {\cal F}^{(1)}_U  {\cal F}^{(1)}_T \right.
\nonumber \\
 & & 
\hspace{4cm}\left.
+(S+\bar{S})(U+\bar{U})\, {\cal F}^{(1)}_S  {\cal F}^{(1)}_U
+(T+\bar{T})(S+\bar{S}) {\cal F}^{(1)}_T  {\cal F}^{(1)}_S
\right]
\nonumber \\
&&
+ \bigg[ \frac{{\cal F}^{(1)}_T}{U + \bar{U}}
+  \frac{{\cal F}^{(1)}_U} {T+\bar{T}} \bigg]
\bigg[ \frac{{\cal F}^{(1)}_U}{S+\bar{S}}
+  \frac{{\cal F}^{(1)}_S}{U + \bar{U}} \bigg]
\bigg[  \frac{{\cal F}^{(1)}_S} {T+\bar{T}}+ \frac{{\cal
      F}^{(1)}_T}{S+\bar{S}} \bigg]  \nonumber \\
&& +  \frac{2 {\cal F}^{(1)}_S \,{\cal F}^{(1)}_T\, {\cal F}^{(1)}_U }
{ (S+\bar{S})(T+\bar{T}) (U+\bar{U}) }\,.
\end{eqnarray}
At large $\modS$, this expression reduces to 
the expression (\ref{nholf3}), where again we made use of (\ref{assum}).


\section{Power-series expansion of $\cFgc{2}$ 
\label{hyperelliptic}}

\setcounter{equation}{0}

In the following, we will consider the power-series expansion of 
$\cFgc{2}$, given in (\ref{fullf2cov}), 
in the limit where ${\bar T} \rightarrow \infty, 
{\bar U} \rightarrow \infty$.  In this limit, $\cFgc{2}$
turns into
\beqa
&&\cFg{2,{\rm top}} (T,U)= \\
&&\hspace{1cm} {\textstyle\frac{1}{4}} a^2
\partial_T^2 \partial_U^2 h 
+ \Big( \frac{a^2}{4 \pi} + \beta_2 \Big)\, \partial_T \partial_U 
\log(j(T) - j(U)) 
- \frac{a\,b_{\rm grav}}{2 \pi} G_2(T) G_2 (U) \label{topf2} \nonumber\\ 
&&\hspace{1cm} + a \Big( \beta_1 - \frac{a}{2 \pi} \Big)\,\Big[G_2(T)\,
\partial_U \log(j(T) - j(U))  +  
G_2(U)\, \partial_T \log(j(T) - j(U)) \Big] \,. \nonumber
\eeqa
In \cite{DWKLL} it was shown that the one-loop correction $h$ to
the prepotential satisfies
\beq
\partial^3_T h(T,U) =  2\,  \frac{E_4(T)\,E_4(U)\, 
E_6(U)\,\eta^{-24}(U)}{j(T)-j(U)}\,, 
\eeq
as well as a similar equation with $T$ and $U$ interchanged. 
The exact expression for $h$ was given in \cite{HarvMoo} in terms
of a power-series expansion. Here we draw attention to the fact
that  this correction carries a different normalization in \cite{DWKLL} 
and \cite{HarvMoo} and was denoted
by $h^{(1)}$, while in this paper $h^{(1)}$ denotes the one-loop correction
to $\cFg{1}$.  Using the result of \cite{HarvMoo}, it follows that 
the first line of (\ref{topf2}) has the following power-series
expansion 
\beqa
\partial_T^2 \partial_U^2 h &=& - 4 \pi\,
\sum_{k,l} \,k^2\, l^2\, c_1(kl) \,\frac{ e^{- 2 \pi ( k T + l U)}}
{(1- e^{- 2 \pi ( k T + l U)})^2} \,\,,
\eeqa
where the integers $k$ and $l$ can take the following values:
either $k=1, l=-1$ or $k>0, l=0$ or $k=0,l>0$ or $k>0,l>0$.
The constants $c_1(n)$ are determined by \cite{HarvMoo}
\beqa
\frac{E_4(T)\, E_6(T)}{\eta^{24}(T)} = \sum_{n= -1}^\infty \,c_1(n)\, q^n \,,
\eeqa
where $q$ and $T$ are related by $q=e^{-2\pi T}$.

The second line of (\ref{topf2}), on the other hand, has
the following power-series expansion in the chamber $T > U$ \cite{HarvMoo}
\beqa
\partial_T \partial_U \log(j(T) - j(U))  = - (2 \pi )^2 \,
\sum_{k,l}\, k\,l\, c(kl)\, \frac{ e^{- 2 \pi ( k T + l U)}}
{(1- e^{- 2 \pi ( k T + l U)})^2} \,,
\eeqa
where the constants $c(n)$ are determined by 
\beqa
j(T) - 744 = \sum_{n = -1}^\infty \, c(n)\, q^n \,\,,
\eeqa
and where the integers $k$ and $l$ can take the same values as
indicated above.

The third line of (\ref{topf2}) can be expanded as follows. Using
\cite{CFILQ} 
\beqa
G_2(T) = - \frac{\pi}{6} \left( 1 - 24 \,\sum_{n=1}^{\infty}\,
\sigma_1(n)\, q^n \right) \,\,,
\eeqa
where $\s_1(n)$ denotes the sum of the divisors of $n$, one finds that
\beq
G_2(T)\, G_2(U) = \frac{\pi^2}{36}
\Big( 1 - 24 \sum_{m=1}^{\infty} \sigma_1(m)\,
q_1^m(1+q_3^m)  + 576 \sum_{n,m=1}^{\infty} \sigma_1(m)\,\sigma_1(n)\,
q_1^{n+m}\,q_3^{n} \Big) ,
\label{g2g2}
\eeq
where $q_i = e^{- 2 \pi t_i}$, and where
$t_1 = U, t_3 = T-U$.

A closer look at the fourth line in (\ref{topf2}) shows that
it is regular as $T \rightarrow U$, that is as $q_3 \rightarrow 1$.
Using that
\beqa
\partial_T 
\log(j(T) - j(U)) &=& 2 \pi \bigg( 1 + 
\frac{ e^{- 2 \pi (  T -  U)}}
{1- e^{- 2 \pi (  T -  U)}}
+ \sum_{k,l > 0} k \, c(kl)\,
\frac{ e^{- 2 \pi ( k T + l U)}}
{1- e^{- 2 \pi ( k T + l U)}} \bigg) \nonumber\\
\partial_U \log(j(T) - j(U)) &=& 2 \pi \bigg( -  
\frac{ e^{- 2 \pi (  T -  U)}}
{1- e^{- 2 \pi ( T -  U)}}
+ \sum_{k,l>0} l \, c(kl) \,
\frac{ e^{- 2 \pi ( k T + l U)}}
{1- e^{- 2 \pi ( k T + l U)}} \bigg) \,,
\eeqa
it follows that 
\beqa
&&
G_2(T)\, \partial_U \log(j(T) - j(U))  + 
G_2(U)\, \partial_T \log(j(T) - j(U)) =\nonumber \\ 
&&2 \pi \bigg( G_2(U)   + \frac{q_3}{1-q_3} (G_2(U) - G_2(T)) \\
&&\hspace{8mm} + G_2(U) 
\sum_{k,l > 0} k \, c(kl)
\frac{ q_1^{k+l} q_3^k }
{1- q_1^{k+l} q_3^k }  + G_2(T) 
\sum_{k,l > 0} l \, c(kl)
\frac{ q_1^{k+l} q_3^k }
{1- q_1^{k+l} q_3^k }  \bigg)\,.  \nonumber
\eeqa
First evaluate
\beqa
\frac{q_3}{1-q_3} (G_2(U) - G_2(T)) &=& 4 \pi
\frac{q_3}{1-q_3} \sum_{n=1}^{\infty} \sigma_1(n) \,(q_1^n - q_3^n q_1^n) 
\nonumber\\
&=&  
4 \pi \sum_{n=1}^{\infty}
\sigma_1(n) \,q_1^n \,(q_3 + q_3^2 + \cdots +q_3^n) \,.
\eeqa
Then, it follows that 
\beqa
G_2(U)  
+ \frac{q_3}{1-q_3} (G_2(U) - G_2(T)) 
= - \frac{\pi}{6} \Big( 1 -24 \sum_{n=1}^{\infty} \sum_{s=0}^{n}
\sigma_1(n)\, q_1^n\, q_3^s \Big) \,.
\eeqa
And finally, using that 
\beqa
\frac{ 1}
{1- q_1^{k+l} q_3^k }  = \sum_{m=0}^{\infty}\,  q_1^{m(k+l)} \,q_3^{mk} \,,
\eeqa
one finds that 
\beqa
&& G_2(U) 
\sum_{k,l > 0} k \, c(kl)\,
\frac{ q_1^{k+l}\, q_3^k }
{1- q_1^{k+l} q_3^k }  
+ G_2(T) 
\sum_{k,l > 0} l \, c(kl)\,
\frac{ q_1^{k+l}\, q_3^k }
{1- q_1^{k+l} q_3^k }=  \nonumber\\
&& -\frac{\pi}{6} 
\sum_{k,l>0} \sum_{m=0}^{\infty} (-)^m  
\bigg[ (k+l) \,c(kl) \,q_1^{(m+1)(k+l)}\, q_3^{(m+1)k}
\nonumber\\
&&\hspace{32mm}
- 24 \sum_{n=1}^{\infty} k\, c(kl)\, \sigma_1(n)
\,q_1^{(m+1)(k+l)+n}\, q_3^{(m+1)k} 
\nonumber\\
&&\hspace{32mm}
- 24 \sum_{n=1}^{\infty} l \,
 c(kl)\, \sigma_1(n)\, q_1^{(m+1)(k+l)+n} \,q_3^{(m+1)k+n}
\bigg] \,.
\eeqa

Thus, it follows that $\cFg{2,{\rm top}}$ has the following 
power-series expansion
\beqa
\cFg{2,{\rm top}} &= &\pi \bigg[ - a^2 
\sum_{k,l} k^2\, l^2\, c_1(kl)\, \frac{ q_1^{k+l} \,q_3^k}
{(1-  q_1^{k+l} q_3^k)^2} \\
&&\hspace{5mm}
- (a^2 + 4 \pi \beta_2)\,
\sum_{k,l} k\,l\, c(kl) \frac{ q_1^{k+l} \,q_3^k }
{(1-  q_1^{k+l} q_3^k )^2} \nonumber\\
&&\hspace{5mm}
- \frac{a \,b_{\rm grav}}{72} 
\Big( 1 - 24 \sum_{m=1}^{\infty} \sigma_1(m)\,
q_1^m(1+q_3^m)  + 576 \sum_{n,m=1}^{\infty} \sigma_1(m)\,\sigma_1(n)\,
q_1^{n+m}\,q_3^{n} \Big)\bigg] \nonumber\\
&&
- \frac{\pi a (2 \pi \beta_1 - a )}{6}  
 \bigg[ 1 -24 \sum_{n=1}^{\infty} \sum_{s=0}^{n}
\sigma_1(n)\, q_1^n \,q_3^s 
\nonumber\\
&&\hspace{27mm}
+ 
\sum_{k,l>0} \sum_{m=0}^{\infty} (-)^m  
\Big[ (k+l) \,c(kl) \,q_1^{(m+1)(k+l)}\, q_3^{(m+1)k}\nonumber\\
&&\hspace{42mm}
- 24 \sum_{n=1}^{\infty} k\, c(kl)\, \sigma_1(n)
\,q_1^{(m+1)(k+l)+n}\, q_3^{(m+1)k}  \nonumber\\
&&\hspace{42mm}
- 24 \sum_{n=1}^{\infty} l \,
 c(kl) \,\sigma_1(n) \,q_1^{(m+1)(k+l)+n} \,q_3^{(m+1)k+n}
\Big] \bigg]  \,.\nonumber
\eeqa
Thus, we see that the instanton expansion of $\cFg{2,{\rm top}}$
is determined in terms of known coefficients $c_1(n)$, 
$c(n)$ and $\sigma_1(n)$.


\end{document}